\documentclass{article}
\usepackage{amsmath,amssymb,theorem,mathpazo,color,url}
\usepackage{graphicx}
\usepackage{here}
\allowdisplaybreaks
\newtheorem{thm}{Theorem}[section]
\newtheorem{cor}[thm]{Corollary}
\newtheorem{lem}[thm]{Lemma}
\newtheorem{prop}[thm]{Proposition}
\newtheorem{defn}[thm]{Definition}
\newtheorem{rem}[thm]{Remark}
\newtheorem{ass}[thm]{Assumption}

\usepackage[hang,small,bf]{caption}
\usepackage[subrefformat=parens]{subcaption}
\def\proof {\noindent{\it Proof.}$\quad$$\quad$}
\def\fin   {\hfill{$\Box$}}
\def\l     {\left}
\def\r     {\right}
\def\<     {\langle}
\def\>     {\rangle}

\def\calB  {{\cal B}}

\def\calF  {{\cal F}}

\def\bbC   {{\mathbb C}}
\def\bbD   {{\mathbb D}}
\def\bbE   {{\mathbb E}}
\def\bbF   {{\mathbb F}}
\def\bbL   {{\mathbb L}}
\def\bbN   {{\mathbb N}}
\def\bbP   {{\mathbb P}}

\def\bbR   {{\mathbb R}}
\def\ve    {\varepsilon}
\def\vp    {\varphi}
\def\vt    {\vartheta}

\def\tP    {\bbP^*}
\def\tN    {\widetilde{N}}
\def\theequation{\thesection.\arabic{equation}}
\begin{document}
\title{Local risk-minimization for Barndorff-Nielsen and Shephard models}
\author{Takuji Arai\footnote{Department of Economics, Keio University, e-mail:arai@econ.keio.ac.jp},
        Yuto Imai\footnote{Department of Mathematics, Waseda University}
        and Ryoichi Suzuki\footnote{Department of Mathematics, Keio University}}

\maketitle

\begin{abstract}
We obtain explicit representations of locally risk-minimizing strategies of call and put options for the Barndorff-Nielsen and Shephard models,
which are Ornstein--Uhlenbeck-type stochastic volatility models.
Using Malliavin calculus for L\'evy processes, Arai and Suzuki \cite{AS} obtained a formula for locally risk-minimizing strategies for L\'evy markets under many additional conditions.
Supposing mild conditions, we make sure that the Barndorff-Nielsen and Shephard models satisfy all the conditions imposed in \cite{AS}.
Among others, we investigate the Malliavin differentiability of the density of the minimal martingale measure.
Moreover, some numerical experiments for locally risk-minimizing strategies are introduced.
\end{abstract}


\setcounter{equation}{0}
\section{Introduction}
The objective is to obtain explicit representations of locally risk-minimizing (LRM) strategies of call and put options for the Barndorff-Nielsen and Shephard (BNS) models:
Ornstein--Uhlenbeck (OU)-type stochastic volatility models developed by Barndorff--Nielsen and Shephard \cite{BNS1}, \cite{BNS2}.
On the other hand, local risk-minimization is a very well-known quadratic hedging method of contingent claims for incomplete financial markets.
Although its theoretical aspects have been well developed, little is known about its explicit representations.
Accordingly, Arai and Suzuki \cite{AS} have analyzed this problem for L\'evy markets using Malliavin calculus for L\'evy processes.
They gave in Theorem 3.7 of their paper an explicit formula for LRM strategies including some Malliavin derivatives.
Here, L\'evy markets mean models for which the asset price process is described by a solution to the following stochastic differential equation (SDE):
\begin{equation}
\label{eq-AS}
dS_t=S_{t-}\l[\alpha_tdt+\beta_tdW_t+\int_{\bbR\backslash\{0\}}\gamma_{t,z}\tN(dt,dz)\r],\hspace{3mm}S_0>0,
\end{equation}
where $W$ is a $1$-dimensional Brownian motion, $\tN$ a compensated Poisson random measure; and $\alpha$, $\beta$, and $\gamma$ are predictable processes.
If $\alpha$, $\beta$, and $\gamma$ are deterministic, a representation for LRM strategies is given simply under some mild conditions.
Indeed, \cite{AS} calculated explicitly LRM strategies of call options, Asian options, and lookback options for the deterministic coefficient case.
However, according to Theorem 3.7 in \cite{AS}, one needs to impose many additional conditions on models with random coefficients.
Thus, concrete calculations for such models were set aside.

In this paper, we obtain explicit LRM strategies for BNS models, which are popular examples for the random coefficient case.
In particular, various empirical studies confirm that BNS models capture well important stylized features of financial time series.
In a BNS model, the squared volatility process $\sigma^2$ is given by an OU process driven by a subordinator, that is, a nondecreasing L\'evy process.
More precisely, $\sigma^2$ is given as a solution to the following SDE:
\begin{equation}
\label{SDE-sigma}
d\sigma_t^2=-\lambda\sigma_t^2dt+dH_{\lambda t}, \ \ \ \sigma_0^2>0,
\end{equation}
where $\lambda>0$, and $H$ is a subordinator without drift.
Now, the asset price process $S$ of a BNS model is described as
\begin{equation}
\label{eq-S}
S_t=S_0\exp\l\{\int_0^t\l(\mu-\frac{1}{2}\sigma_s^2\r)ds+\int_0^t\sigma_sdW_s+\rho H_{\lambda t}\r\},
\end{equation}
where $S_0>0$, $\rho\leq0$, $\mu\in\bbR$.
Note that the last term $\rho H_{\lambda t}$ accounts for the leverage effect, which is a stylized fact such that the asset price declines at the moment when volatility increases.
Moreover, defining $J_t:=H_{\lambda t}$, we denote by $N$ the Poisson random measure of $J$.
Hence, we have $J_t=\int_0^\infty xN([0,t],dx)$. Denoting by $\nu$ the L\'evy measure of $J$, we find $\tN(dt,dx):=N(dt,dx)-\nu(dx)dt$ is the compensated Poisson random measure.
Then, the asset price process $S$ given in (\ref{eq-S}) is a solution to the following SDE:
\begin{equation}
\label{SDE}
dS_t=S_{t-}\l\{\alpha dt+\sigma_tdW_t+\int_0^\infty(e^{\rho x}-1)\tN(dt,dx)\r\},
\end{equation}
where $\alpha:=\mu+\int_0^\infty(e^{\rho x}-1)\nu(dx)$.
Therefore, BNS models correspond to instances where $\beta$ in (\ref{eq-AS}) is random.

We shall use Theorem 3.7 of \cite{AS} in this paper to derive LRM strategies for BNS models described as in (\ref{eq-S}).
Therefore, the primary part of our discussion lies in confirming all the conditions imposed on Theorem 3.7 of \cite{AS}.
In particular, we need to investigate the Malliavin differentiability of the density of the minimal martingale measure (MMM),
which is an indispensable equivalent martingale measure to discuss LRM strategies.
To the best of our knowledge, literature on LRM strategies for BNS models is very limited.
Arai \cite{A} studied this problem for a different setting from ours.
In \cite{A}, the volatility risk premium is taken into account, but $\rho$ is restricted to $0$.
Hence, $S$ is described as
\[
S_t=S_0\exp\l\{\int_0^t\l(\mu+\beta\sigma_s^2\r)ds+\int_0^t\sigma_sdW_s\r\},
\]
where $\beta\in\bbR$ is called the volatility risk premium.
Note that $S$ is continuous.
Formulating a Malliavin calculus under the MMM, \cite{A} gave an explicit representation of LRM strategies.
On the other hand, there is some previous research on mean-variance hedging, which is an alternative quadratic hedging method, for BNS models.
Cont, Tankov and Voltchkova \cite{CTV}, and Kallsen and Pauwels \cite{KP} studied this problem assuming $S$ is a martingale.
Kallsen and Vierthauer \cite{KV} treated the case where $\rho=0$.
Recently, Benth and Detering \cite{BD} dealt with the BNS model framework to represent a future price process on electricity assuming that $S$ is a martingale and $\rho=0$.

In addition, we also develop in this paper a numerical scheme for LRM strategies for call options using the method of Arai, Imai and Suzuki \cite{AIS},
which is a numerical scheme of LRM strategies for exponential L\'evy models.
Their scheme is based on the so-called Carr--Madan approach \cite{CM}, which is based on the fast Fourier transform (FFT).
Moreover, we compare LRM strategies with the so-called delta-hedging strategies, which are given as the partial derivative of the option price with respect to the asset price.

The outline of this paper is as follows.
After giving preliminaries in Section 2, we address the main results in Section 3.
Theorem \ref{thm-main} gives an explicit representation of LRM strategies for put options.
LRM strategies for call options are provided as its corollary.
A proof of Theorem \ref{thm-main} is discussed in Section 4. Section 5 is devoted to the Malliavin differentiability of the density of the MMM.
Numerical experiments for LRM strategies are illustrated in Section 6. Conclusions are given in Section 7.
The statement of Theorem 3.7 of \cite{AS} for our setting, and some additional calculations, are provided in Appendix.

\setcounter{equation}{0}
\section{Preliminaries}
We consider a financial market model in which only one risky asset and one riskless asset are tradable.
For simplicity, we assume the interest rate to be $0$. Let $T$ be a finite time horizon.
The fluctuation of the risky asset is described as a process $S$ given by (\ref{eq-S}).
We adopt the same mathematical framework as in \cite{AS}.
The structure of the underlying probability space $(\Omega, \calF, \bbP)$ will be discussed in Subsection 2.3 below.
Notice that the Poisson random measure $N$ and the L\'evy measure $\nu$ of $J$ are defined on $[0,T]\times(0,\infty)$ and $(0,\infty)$, respectively, and that
\[
\int_0^\infty(x\wedge1)\nu(dx)<\infty
\]
by Proposition 3.10 of Cont and Tankov \cite{CT}. Let $\nu^H$ be the L\'evy measure of $H$; we then have $\nu(dx)=\lambda\nu^H(dx)$.
Denoting $A_t:=\int_0^tS_{s-}\alpha ds$ and $M_t:=S_t-S_0-A_t$, we have $S_t=S_0+M_t+A_t$, which is the canonical decomposition of $S$.
Further, we denote $L_t:=\log(S_t/S_0)$ for $t\in[0,T]$, that is,
\begin{equation}
\label{eq-L}
L_t=\int_0^t\l(\mu-\frac{1}{2}\sigma_s^2\r)ds+\int_0^t\sigma_sdW_s+\rho J_t.
\end{equation}

\begin{rem}
Noting that $\sigma_{t-}=\sigma_t$ a.s. for any $t\in[0,T]$, we can regard $\sigma_t$ and $\sigma^2_t$ as predictable processes.
For example, we may identify $\sigma_tdW_t$ in (\ref{SDE}) with $\sigma_{t-}dW_t$, if necessary.
\end{rem}

Next, we state our standing assumptions:

\begin{ass}
\label{ass1}
 \
\begin{enumerate}
\item
$\int_1^\infty\exp\{2(\calB(T)\vee|\rho|)x\}\nu(dx)<\infty$, where $\calB(t):=\frac{1-e^{-\lambda t}}{\lambda}$ for $t\in[0,T]$.
\item
$\frac{\alpha}{e^{-\lambda T}\sigma_0^2+C_\rho}>-1$, where $C_\rho:=\int_0^\infty(e^{\rho x}-1)^2\nu(dx)$.
\end{enumerate}
\end{ass}

\begin{rem}
\label{rem1}
 \
\begin{enumerate}
\item
Item 1 in Assumption \ref{ass1} ensures $\int_0^\infty x^2\nu(dx)<\infty$, which means $\bbE[J_T^2]<\infty$.
In addition, we have $\int_0^\infty(e^{\rho x}-1)^2\nu(dx)\leq\int_0^\infty\rho^2x^2\nu(dx)<\infty$, because $0\leq1-e^{\rho x}\leq -\rho x$.
\item
As seen in Subsection 2.3 of \cite{AS}, the so-called (SC) condition is satisfied under Assumption \ref{ass1}.
For more details on the (SC) condition, see Schweizer \cite{Sch}, \cite{Sch3}.
Moreover, Lemma 2.11 of \cite{AS} implies that $\bbE\l[\sup_{t\in[0,T]}|S_t|^2\r]<\infty$.
\item
By (\ref{eq-sigma-1}) in Appendix, item 2 ensures that $\frac{\alpha}{\sigma_t^2+C_\rho}>-1$ for any $t\in[0,T]$.
\end{enumerate}
\end{rem}

\begin{rem}
We state two important examples of $\sigma^2$ introduced in Nicolato and Venardos \cite{NV} that fulfill Assumption 2.2 under certain conditions on the involved parameters.
For more details on this topic, see also Schoutens \cite{Scho}.
\begin{enumerate}
\item
The first concerns $\nu^H$ given by
\[
\nu^H(dx)=\frac{a}{2\sqrt{2\pi}}x^{-\frac{3}{2}}(1+b^2x)e^{-\frac{1}{2}b^2x}{\bf 1}_{(0,\infty)}(x)dx
\]
where $a>0$ and $b>0$. In this case, the invariant distribution of the squared volatility process $\sigma^2$ follows an inverse-Gaussian distribution with parameters $a>0$ and $b>0$.
$\sigma^2$ is called an IG-OU process. If $\frac{b^2}{2}>2(\calB(T)\vee|\rho|)$, then item 1 of Assumption 2.2 is satisfied.
\item
The second example is what we shall call a Gamma-OU process, where the invariant distribution of $\sigma^2$ is given by a Gamma distribution with parameters $a>0$ and $b>0$.
In this case, $\nu^H$ is described as
\[
\nu^H(dx)=abe^{-bx}{\bf 1}_{(0,\infty)}(x)dx.
\]
As well as the IG-OU case, item 1 of Assumption 2.2 is satisfied if $b>2(\calB(T)\vee|\rho|)$.
\item
\cite{NV} and Section 7 in \cite{Scho} estimated the parameter sets for the above two models using real data.

\begin{table}[htb]
\begin{center}
\caption{Estimated parameters for IG-OU and Gamma-OU processes} 
\begin{tabular}{cccccc} \hline
IG-OU       & $\rho$    & $\lambda$ & $a$      & $b$        & $\sigma_0^2$ \\ \hline \hline
\cite{NV}   & $-4.7039$ & $2.4958$  & $0.0872$ & $11.9800$  & $0.0041$     \\ \hline
\cite{Scho} & $-0.1926$ & $0.0636$  & $6.2410$ & $0.7995$   & $0.0156$     \\ \hline \hline
Gamma-OU    &           &           &          &            &              \\ \hline
\cite{NV}   & $-4.4617$ & $1.6787$  & $1.0071$ & $116.0100$ & $0.0043$     \\ \hline
\cite{Scho} & $-1.2606$ & $0.5783$  & $1.4338$ & $11.6641$  & $0.0145$     \\ \hline
\end{tabular}
\end{center}
\end{table}

\vspace{-5mm}

\noindent
Note that the discounted asset price process is assumed to be a martingale in both \cite{NV} and \cite{Scho}.
Hence, the value of $\mu$ is automatically determined. For any $T>0$, the parameter set for IG-OU in \cite{Scho} does not satisfy item 1 of Assumption 2.2.
In contradistinction, the other estimated parameter sets listed in Table 1 satisfy the condition.
\end{enumerate}
\end{rem}

\subsection{Locally risk-minimizing strategies}

In this subsection, we give a definition of LRM strategies based on Theorem 1.6 of \cite{Sch3}.

\begin{defn}
 \
\begin{enumerate}
\item
$\Theta_S$ denotes the space of all $\bbR$-valued predictable processes $\xi$
satisfying $\bbE\l[\int_0^T\xi_t^2d\langle M\rangle_t+(\int_0^T|\xi_tdA_t|)^2\r]<\infty$.
\item
An $L^2$-strategy is given by a pair $\vp=(\xi,\eta)$,
where $\xi\in\Theta_S$ and $\eta$ is an adapted process such that $V(\vp):=\xi S+\eta$ is a right continuous process with $\bbE[V_t^2(\vp)]<\infty$ for every $t\in[0,T]$.
Note that $\xi_t$ (resp. $\eta_t$) represents the number of units of the risky asset (resp., the risk-free asset) an investor holds at time $t$.
\item
For claim $F\in L^2(\bbP)$, the process $C^F(\vp)$ defined by $C^F_t(\vp):=F1_{\{t=T\}}+V_t(\vp)-\int_0^t\xi_s dS_s$ is called the cost process of $\vp=(\xi, \eta)$ for $F$.
\item
An $L^2$-strategy $\vp$ is said to be locally risk-minimizing (LRM) for claim $F$ if $V_T(\vp)=0$ and $C^F(\vp)$ is a martingale orthogonal to $M$, that is,
$[C^F(\vp),M]$ is a uniformly integrable martingale.
\item
An $F\in L^2(\bbP)$ admits a F\"ollmer--Schweizer (FS) decomposition if it can be described by
\begin{equation}
\label{eqFS}
F=F_0+\int_0^T\xi_t^FdS_t+L_T^F,
\end{equation}
where $F_0\in\bbR$, $\xi^F\in\Theta_S$ and $L^F$ is a square-integrable martingale orthogonal to $M$ with $L_0^F=0$.
\end{enumerate}
\end{defn}

\noindent
For more details on LRM strategies, see \cite{Sch}, \cite{Sch3}.
We now introduce Proposition 5.2 of \cite{Sch3}.

\begin{prop}[Proposition 5.2 of \cite{Sch3}]
Under Assumption \ref{ass1}, an LRM strategy $\vp=(\xi,\eta)$ for $F$ exists if and only if $F$ admits an FS decomposition; and its relationship is given by
\[
\xi_t=\xi^F_t,\hspace{3mm}\eta_t=F_0+\int_0^t\xi^F_sdS_s+L^F_t-F1_{\{t=T\}}-\xi^F_tS_t.
\]
\end{prop}

\noindent
Therefore, it suffices to derive a representation of $\xi^F$ in (\ref{eqFS}) to obtain the LRM strategy for claim $F$.
Henceforth, we identify $\xi^F$ with the LRM strategy for $F$.

\subsection{Minimal martingale measure}

To discuss the FS decomposition, we first need to study the MMM. A probability measure $\tP\sim\bbP$ is called an MMM,
if $S$ is a $\tP$-martingale; and any square-integrable $\bbP$-martingale orthogonal to $M$ remains a martingale under $\tP$.
Next we consider the following SDE:
\begin{equation}
\label{SDE-Z}
dZ_t=-Z_{t-}\Lambda_tdM_t, \ \ \ Z_0=1,
\end{equation}
where $\Lambda_t:=\frac{1}{S_{t-}}\frac{\alpha}{\sigma_t^2+C_\rho}$.
The solution to (\ref{SDE-Z}) is a stochastic exponential of $-\int_0^\cdot\Lambda_tdM_t$.
More precisely, denoting
\[
u_s:=\Lambda_sS_{s-}\sigma_s=\frac{\alpha\sigma_s}{\sigma_s^2+C_\rho}\qquad\mbox{and}\quad\theta_{s,x}:=\Lambda_sS_{s-}(e^{\rho x}-1)=\frac{\alpha(e^{\rho x}-1)}{\sigma_s^2+C_\rho}
\]
for $s\in[0,T]$ and $x\in(0,\infty)$, we have $\Lambda_tdM_t=u_tdW_t +\int_0^\infty\theta_{t,z}\tN(dt,dz)$; and
\begin{align}
Z_t &= \exp\bigg\{-\int_0^tu_sdW_s-\frac{1}{2}\int_0^tu_s^2ds+\int_0^t\int_0^\infty\log(1-\theta_{s,x})\tN(ds,dx) \nonumber \\
    &  \hspace{5mm}+\int_0^t\int_0^\infty(\log(1-\theta_{s,x})+\theta_{s,x})\nu(dx)ds\bigg\}.
\label{eq-MMM-1}
\end{align}
We remark here that
\[
\int_0^T\int_0^\infty\l\{|\log(1-\theta_{s,x})|^2+\theta_{s,x}^2\r\}\nu(dx)ds \leq 2TC_\theta^2\rho^2\int_0^\infty x^2\nu(dx) < \infty
\]
by Lemma \ref{lem-ut-1}.
Noting the boundedness of $u_s$ by Lemma \ref{lem-ut-1}, and
\[
(1-\theta_{s,x})\log(1-\theta_{s,x})+\theta_{s,x} \leq (1-\theta_{s,x})(-\theta_{s,x})+\theta_{s,x} = \theta^2_{s,x},
\]
we have the martingale property of $Z$ by Theorem 1.4 of Ishikawa \cite{I}.
Now, we get the following:

\begin{prop}
\label{prop-MMM}
 \
\begin{enumerate}
\item $Z_T\in L^2(\bbP)$.
\item The probability measure defined by $\frac{d\tP}{d\bbP}=Z_T$ is the MMM.
\end{enumerate}
\end{prop}

\proof
We first demonstrate item 1. Here (\ref{eq-MMM-1}) and Lemma \ref{lem-ut-1} imply that
\begin{align*}
Z_T^2
&=    \exp\bigg\{-\int_0^T2u_sdW_s-\frac{1}{2}\int_0^T4u_s^2ds+\int_0^T\int_0^\infty\log(1-\delta_{s,x})\tN(ds,dx) \\
&     \hspace{5mm}+\int_0^T\int_0^\infty\l[\log(1-\delta_{s,x})+\delta_{s,x}+\theta_{s,x}^2\r]\nu(dx)ds+\int_0^Tu_s^2ds\bigg\} \\
&\leq \exp\bigg\{-\int_0^T2u_sdW_s-\frac{1}{2}\int_0^T4u_s^2ds+\int_0^T\int_0^\infty\log(1-\delta_{s,x})\tN(ds,dx) \\
&     \hspace{5mm}+\int_0^T\int_0^\infty\l[\log(1-\delta_{s,x})+\delta_{s,x}\r]\nu(dx)ds+T(C_\theta^2C_\rho+C_u^2)\bigg\}
\end{align*}
where $\delta_{s,x}:=2\theta_{s,x}-\theta_{s,x}^2$, and $C_u$ and $C_\theta$ are constants defined in (\ref{eq-def-C}). That is, denoting
\begin{align}
Y_t &:= \exp\bigg\{-\int_0^t2u_sdW_s-\frac{1}{2}\int_0^t4u_s^2ds+\int_0^t\int_0^\infty\log(1-\delta_{s,x})\tN(ds,dx) \nonumber \\
    &   \hspace{5mm}+\int_0^t\int_0^\infty\l[\log(1-\delta_{s,x})+\delta_{s,x}\r]\nu(dx)ds\bigg\}
\label{eq-MMM-2}
\end{align}
for $t\in[0,T]$, we have
\begin{equation}
\label{eq-MMM-3}
Z^2_T\leq Y_T\exp\{T(C_\theta^2C_\rho+C_u^2)\}.
\end{equation}
Therefore, we need only to show the process $Y$ is a martingale.
First, the Brownian part of $Y$ is a martingale as $u_s$ is bounded. Lemma \ref{lem-ut-1} again yields
\[
\int_0^T\int_0^\infty|\log(1-\delta_{s,x})|^2\nu(dx)ds \leq \int_0^T\int_0^\infty4C_\theta^2\rho^2x^2\nu(dx)ds < \infty;
\]
and $\delta_{s,x}^2=\theta_{s,x}^2(2-\theta_{s,x})^2\leq C_\theta^2\rho^2x^2 (2+C_\theta)^2$, that is, $\int_0^T\int_0^\infty\delta_{s,x}^2\nu(dx)ds<\infty$.
In addition, we have
\[
\int_0^T\int_0^\infty[(1-\delta_{s,x})\log(1-\delta_{s,x})+\delta_{s,x}]\nu(dx)ds \leq \int_0^T\int_0^\infty\delta^2_{s,x}\nu(dx)ds < \infty.
\]
Hence, all the conditions in Theorem 1.4 of \cite{I} are satisfied, that is, $Y$ is a martingale.

We proceed to item 2. The martingale property of $Z$ implies that the product process $ZS$ is a $\bbP$-local martingale.
Thus, $S$ is a $\tP$-martingale, because $\sup_{t\in[0,T]}|S_t|$ and $Z_T$ are in $L^2(\bbP)$.
Moreover, letting $L$ be a square-integrable $\bbP$-martingale with null at $0$ orthogonal to $M$, we have that $LZ$ is a $\bbP$-local martingale.
By the square integrability of $L$, $L$ remains a martingale under $\tP$.
Therefore, $\tP$ is the MMM. This completes the proof of Proposition \ref{prop-MMM}.
\fin

\subsection{Malliavin calculus}

In this subsection, we prepare Malliavin calculus based on the canonical L\'evy space framework undertaken by Sol\'e, Utzet and Vives \cite{S07}.
The underlying probability space $(\Omega, \calF, \bbP)$ is assumed to be given by $(\Omega_W\times\Omega_J, \calF_W\times\calF_J, \bbP_W\times\bbP_J)$,
where $(\Omega_W, \calF_W, \bbP_W)$ is a $1$-dimensional Wiener space on $[0,T]$ with coordinate mapping process $W$;
and $(\Omega_J, \calF_J, \bbP_J)$ is the canonical L\'evy space for $J$, that is, $\Omega_J=\cup_{n=0}^\infty([0,T]\times(0,\infty))^n$;
and $J_t(\omega_J)=\sum_{i=1}^nz_i{\bf 1}_{\{t_i\leq t\}}$ for $t\in[0,T]$ and $\omega_J=((t_1,z_1),\dots,(t_n,z_n)) \in([0,T]\times(0,\infty))^n$.
Note that $([0,T]\times(0,\infty))^0$ represents an empty sequence. Let $\bbF=\{\calF_t\}_{t\in[0,T]}$ be the canonical filtration completed for $\bbP$.
For more details, see Delong and Imkeller \cite{DI}, and \cite{S07}.

To begin, we define measures $q$ and $Q$ on $[0,T]\times[0,\infty)$ as
\[
q(E):=\int_E\delta_0(dz)dt+\int_Ez^2\nu(dz)dt,
\]
and
\[
Q(E):=\int_E\delta_0(dz)dW_t+\int_Ez\tN(dt,dz),
\]
where $E\in\calB([0,T]\times[0,\infty))$ and $\delta_0$ is the Dirac measure at $0$.
For $n\in\bbN$, we denote by $L_{T,q,n}^2$ the set of product measurable, deterministic functions $h:([0,T]\times[0,\infty))^n\to\bbR$ satisfying
\[
\|h\|_{ L_{T,q,n}^2}^2:=\int_{([0,T]\times[0,\infty))^n}|h((t_1,z_1),\cdots,(t_n,z_n))|^2q(dt_1,dz_1)\cdots q(dt_n,dz_n)<\infty.
\]
For $n\in\bbN$ and $h\in L_{T,q,n}^2$, we define
\[
I_n(h):=\int_{([0, T]\times[0,\infty))^n}h((t_1,z_1),\cdots,(t_n,z_n))Q(dt_1,dz_1)\cdots Q(dt_n,dz_n).
\]
Formally, we denote $L_{T,q,0}^2:=\bbR$ and $I_0(h):=h$ for $h\in\bbR$.
Under this setting, any $F\in L^2(\bbP)$ has the unique representation $F=\sum_{n=0}^{\infty}I_n(h_n)$ with functions $h_n\in L_{T,q,n}^2$ that are symmetric in the $n$ pairs
$(t_i,z_i), 1\leq i\leq n$, and we have $\bbE[F^2]=\sum_{n=0}^\infty n!\|h_n\|_{L_{T,q,n}^2}^2$.
We define a Malliavin derivative operator.

\begin{defn}
\label{def-Malliavin}
 \
\begin{enumerate}
\item
Let $\bbD^{1,2}$ denote the set of $\calF$-measurable random variables $F\in L^2(\bbP)$ with $F=\sum_{n=0}^\infty I_n(h_n)$ satisfying $\sum_{n=1}^\infty nn!\|h_n\|_{L_{T,q,n}^2}^2<\infty$.
\item
For any $F\in\bbD^{1,2}$, a Malliavin derivative $DF:[0,T]\times[0,\infty)\times\Omega\to\bbR$ is defined as
\[
D_{t,z}F=\sum_{n=1}^\infty nI_{n-1}(h_n((t,z),\cdot))
\]
for $q$-a.e. $(t,z)\in[0,T]\times[0,\infty)$, $\bbP$-a.s.
\end{enumerate}
\end{defn}

\setcounter{equation}{0}
\section{Main results}
Using the framework of Theorem 3.7 of \cite{AS}, we introduce in this section explicit representations of LRM strategies for call and put options as the main results of this paper.
Note that the statement of Theorem 3.7 of \cite{AS} for our setting is introduced in Appendix as Theorem \ref{thm-AS37}.
To this end, denoting by $F$ the underlying contingent claim, we need $Z_TF\in L^2(\bbP)$ (Condition AS1 in Theorem \ref{thm-AS37}).
If $F$ is a call option, this condition is not necessarily satisfied in our setting.
On the other hand, because put options are bounded, we need not care about any integrability condition for them.
Therefore, we treat put options first and derive LRM strategies for call options from the put--call parity.
With this procedure, we can do without any additional assumptions.

\begin{thm}
\label{thm-main}
For $K>0$, the LRM strategy $\xi^{(K-S_T)^+}$ of put option $(K-S_T)^+$ is represented as
\begin{align}
\xi_t^{(K-S_T)^+}
&= \frac{1}{S_{t-}(\sigma_t^2+C_\rho)}\bigg\{\sigma_t^2\bbE_{\bbP^*}[-{\bf 1}_{\{S_T<K\}}S_T|\calF_{t-}]\nonumber \\
&  \hspace{5mm}+\int_0^\infty\bbE_{\tP}[(K-S_T)^+(H^*_{t,z}-1)+zH^*_{t,z}D_{t,z}(K-S_T)^+|\calF_{t-}](e^{\rho z}-1)\nu(dz)\bigg\},
\label{eq-thm-main}
\end{align}
where $D_{t,z}(K-S_T)^+$ is given by Proposition \ref{prop-put-deriv}; and
\[
H^*_{t,z}:= \exp\{zD_{t,z}\log Z_T-\log(1-\theta_{t,z})\}
\]
for $(t,z)\in[0,T]\times(0,\infty)$.
Note that $D_{t,z}\log Z_T$ is provided in Proposition \ref{prop-logZT}.
\end{thm}

\begin{rem}
To obtain a more explicit representation of $\xi_t^{(K-S_T)^+}$, we calculate the conditional expectation in the second term of (\ref{eq-thm-main}) as, for $z\in(0,\infty)$,
\begin{align*}
\lefteqn{\bbE_{\tP}[(K-S_T)^+(H^*_{t,z}-1)+zH^*_{t,z}D_{t,z}(K-S_T)^+|\calF_{t-}]} \\
&= \bbE_{\tP}[H^*_{t,z}\{(K-S_T)^++zD_{t,z}(K-S_T)^+\}-(K-S_T)^+|\calF_{t-}] \\
&= \frac{\bbE[Z_TH^*_{t,z}\{(K-S_T)^++zD_{t,z}(K-S_T)^+\}|\calF_{t-}]}{Z_{t-}}-\bbE_{\tP}[(K-S_T)^+|\calF_{t-}] \\
&= \frac{\bbE[Z_TH^*_{t,z}(K-S_T\exp\{zD_{t,z}L_T\})^+|\calF_{t-}]}{Z_{t-}}-\bbE_{\tP}[(K-S_T)^+|\calF_{t-}],
\end{align*}
where $D_{t,z}L_T$ is given explicitly by Proposition \ref{prop-LT-1}. Note that the last equality is implied by Proposition \ref{prop-put-deriv} below.

We now calculate $\frac{Z_TH^*_{t,z}}{Z_{t-}}$, and investigate its properties for later use. For $t\in[0,T]$, $z\in(0,\infty)$, $s\in[t,T]$, and $x\in(0,\infty)$, we denote
\[
A^u_{t,z,s} := u_s+zD_{t,z}u_s = f_u\l(\sqrt{\sigma^2+ze^{-\lambda(s-t)}}\r) = \frac{\alpha\sqrt{\sigma^2_s+ze^{-\lambda(s-t)}}}{\sigma^2_s+ze^{-\lambda(s-t)}+C_\rho},
\]
and
\begin{equation}
\label{eq-thm-main-rem-2}
A^\theta_{t,z,s,x} := \theta_{s,x}+zD_{t,z}\theta_{s,x}
                    = f_\theta\l(\sqrt{\sigma^2_s+ze^{-\lambda(s-t)}}\r)(e^{\rho x}-1)
                    = \frac{\alpha(e^{\rho x}-1)}{\sigma^2_s+ze^{-\lambda(s-t)}+C_\rho}
\end{equation}
by Lemmas \ref{lem-ut-2} and \ref{lem-ut-3}.
We obtain then, by (\ref{eq-MMM-1}), Lemmas \ref{lem-ut-2}--\ref{lem-ut-4}, and
Proposition \ref{prop-logZT},
\begin{align*}
\frac{Z_TH^*_{t,z}}{Z_{t-}}
&= \exp\bigg\{-\int_t^T(u_s+zD_{t,z}u_s)dW_s-\frac{1}{2}\int_t^T(u_s+zD_{t,z}u_s)^2ds \\
&  \hspace{7mm}+\int_{t-}^T\int_0^\infty\l[\log(1-\theta_{s,x})+zD_{t,z}\log(1-\theta_{s,x})\r]\tN(ds,dx) \\
&  \hspace{7mm}+\int_t^T\int_0^\infty\l[\log(1-\theta_{s,x})+zD_{t,z}\log(1-\theta_{s,x})+\theta_{s,x}+zD_{t,z}\theta_{s,x}\r]\nu(dx)ds\bigg\} \\
&= \exp\bigg\{-\int_t^TA^u_{t,z,s}dW_s-\frac{1}{2}\int_t^T(A^u_{t,z,s})^2ds \\
&  \hspace{7mm}+\int_{t-}^T\int_0^\infty\log(1-A^\theta_{t,z,s,x})\tN(ds,dx) \\
&  \hspace{7mm}+\int_t^T\int_0^\infty\l[\log(1-A^\theta_{t,z,s,x})+A^\theta_{t,z,s,x}\r]\nu(dx)ds\bigg\}.
\end{align*}
Note that $A^u_{t,z,s}$ is bounded.
Moreover, (\ref{eq-thm-main-rem-2}) and (\ref{eq-ut-5}) imply that $\int_0^\infty (A^\theta_{t,z,s,x})^2\nu(dx) <C^2_\theta C_\rho$ and $A^\theta_{t,z,s,x}\leq1-e^{\rho x}$.
We have then
\[
|\log(1-A^\theta_{t,z,s,x})|^2
\leq\l\{
    \begin{array}{ll}
       \rho^2x^2,              & \mbox{ if }A^\theta_{t,z,s,x}>0, \\
       (A^\theta_{t,z,s,x})^2, & \mbox{ otherwise},
    \end{array}\r.
\]
which implies that $\int_0^\infty|\log(1-A^\theta_{t,z,s,x})|^2\nu(dx)<\infty$. As a result, we have
\begin{equation}
\label{eq-thm-main-rem-3}
\bbE\l[\frac{Z_TH^*_{t,z}}{Z_{t-}}\bigg|\calF_{t-}\r]=1
\end{equation}
from the view of Theorem 1.4 in \cite{I}.
\end{rem}

\begin{cor}
\label{cor1}
The LRM strategy for call option $(S_T-K)^+$ is given as $\xi^{(S_T-K)^+}=1+\xi^{(K-S_T)^+}$.
\end{cor}

\proof
Note that $S$ is a $\tP$-martingale by Remark \ref{rem1} and Proposition \ref{prop-MMM}.
We then obtain
\begin{align*}
(S_T-K)^+
&= S_T-K+(K-S_T)^+ \\
&= S_0+\int_0^TdS_t-K+\bbE_{\bbP^*}[(K-S_T)^+]+\int_0^T\xi^{(K-S_T)^+}_tdS_t+L^{(K-S_T)^+}_T \\
&= \bbE_{\bbP^*}\l[S_T-K+(K-S_T)^+\r]+\int_0^T\l(1+\xi^{(K-S_T)^+}_t\r)dS_t+L^{(K-S_T)^+}_T \\
&= \bbE_{\bbP^*}\l[(S_T-K)^+\r]+\int_0^T\l(1+\xi^{(K-S_T)^+}_t\r)dS_t+L^{(K-S_T)^+}_T,
\end{align*}
where $L^{(K-S_T)^+}$ is defined in (\ref{eqFS}).
This is an FS decomposition of $(S_T-K)^+$ as $1\in\Theta_S$ by the (SC) condition.
\fin

\setcounter{equation}{0}
\section{Proof of Theorem \ref{thm-main}}
\label{sect-proof}
We begin with the Malliavin derivatives of put options.

\begin{prop}
\label{prop-put-deriv}
For $K>0$, we have $(K-S_T)^+\in\bbD^{1,2}$ and
\begin{align*}
D_{t,z}(K-S_T)^+
&= -{\bf 1}_{\{S_T<K\}}S_TD_{t,0}L_T\cdot{\bf 1}_{\{0\}}(z) \\
&  \hspace{5mm}+\frac{(K-S_Te^{zD_{t,z}L_T})^+-(K-S_T)^+}{z}{\bf 1}_{(0,\infty)}(z).
\end{align*}
\end{prop}

\proof
First, note that $S_T=S_0e^{L_T}$, and $L_T\in\bbD^{1,2}$ by Proposition \ref{prop-LT-1}. However, $S_T$ is not necessarily Malliavin differentiable.
Hence, we regard $(K-S_T)^+$ as a functional of $L_T$ rather than $S_T$ to calculate its Malliavin derivative.
To this end, noting the boundedness of $(K-S_T)^+$, we introduce the following function:
\[
f_K(r):=
\begin{cases}
S_0e^r,              & \mbox{if } r\leq\log(K/S_0), \\
Kr+K(1-\log(K/S_0)), & \mbox{if } r>\log(K/S_0).
\end{cases}
\]
Then, $f_K\in C^1(\bbR)$ and $0<f^\prime_K(r)\leq K$ for any $r\in\bbR$. We also note $(K-S_T)^+=(K-f_K(L_T))^+$.
Proposition 2.6 in \cite{Suz} implies that $f_K(L_T)\in\bbD^{1,2}$ and
\[
D_{t,z}f_K(L_T)=f^\prime_K(L_T)D_{t,0}L_T\cdot{\bf 1}_{\{0\}}(z)+\frac{f_K(L_T+zD_{t,z}L_T)-f_K(L_T)}{z}{\bf 1}_{(0,\infty)}(z).
\]
The same argument as Theorem 4.1 of \cite{AS} implies that, for $q$-a.e. $(t,z)\in[0,T]\times[0,\infty)$,
\begin{align*}
D_{t,z}(K-S_T)^+
&= D_{t,z}(K-f_K(L_T))^+ \\
&= -{\bf 1}_{\{f_K(L_T)<K\}}D_{t,0}f_K(L_T)\cdot{\bf 1}_{\{0\}}(z) \\
&  \hspace{5mm}+\frac{(K-f_K(L_T)-zD_{t,z}f_K(L_T))^+-(K-f_K(L_T))^+}{z}{\bf 1}_{(0,\infty)}(z) \\
&= -{\bf 1}_{\{S_T<K\}}S_TD_{t,0}L_T\cdot{\bf 1}_{\{0\}}(z) \\
&  \hspace{5mm}+\frac{(K-f_K(L_T+zD_{t,z}L_T))^+-(K-f_K(L_T))^+}{z}{\bf 1}_{(0,\infty)}(z) \\
&= -{\bf 1}_{\{S_T<K\}}S_TD_{t,0}L_T\cdot{\bf 1}_{\{0\}}(z) \\
&  \hspace{5mm}+\frac{(K-S_Te^{zD_{t,z}L_T})^+-(K-S_T)^+}{z}{\bf 1}_{(0,\infty)}(z).
\end{align*}
\fin

We now prove Theorem \ref{thm-main} through Theorem \ref{thm-AS37} (Theorem 3.7 of \cite{AS}).
To this end, we need only to make sure of Conditions AS2 and AS3 in Theorem \ref{thm-AS37}.
Note that Condition AS1 is ensured by Proposition \ref{prop-MMM} and the boundedness of $(K-S_T)^+=:F$.
We first confirm Condition AS2 listed below:

\begin{description}
\item[C1]
$u$, $u^2\in\bbL^{1,2}_0$; and $2u_sD_{t,z}u_s+z(D_{t,z}u_s)^2\in L^2(q\times\bbP)$ for a.e. $s\in[0,T]$.
\item[C2]
$\theta+\log(1-\theta)\in\widetilde{\bbL}_1^{1,2}$, and $\log(1-\theta)\in\bbL_1^{1,2}$.
\item[C3]
For $q$-a.e. $(s,x)\in[0,T]\times(0,\infty)$, there is an $\ve_{s,x}\in(0,1)$ such that $\theta_{s,x}<1-\ve_{s,x}$.
\item[C4]
$Z_T\l\{D_{t,0}\log Z_T{\bf 1}_{\{0\}}(z)+\frac{e^{zD_{t,z}\log Z_T}-1}{z}{\bf 1}_{(0,\infty)}(z)\r\}\in L^2(q\times\bbP)$.
\item[C5]
$F\in\bbD^{1,2}$; and $Z_TD_{t,z}F+FD_{t,z}Z_T+zD_{t,z}F\cdot D_{t,z}Z_T\in L^2(q\times\bbP)$.
\item[C6]
$FH^*_{t,z}$, $H^*_{t,z}D_{t,z}F\in L^1(\tP)$ for $q$-a.e. $(t,z)\in[0,T]\times(0,\infty)$.
\end{description}
Here $\bbL_0^{1,2}$, $\bbL_1^{1,2}$ and $\widetilde{\bbL}_1^{1,2}$ are defined as follows:
\begin{itemize}
\item
$\bbL_0^{1,2}$ denotes the space of $G:[0,T]\times\Omega\to\bbR$ satisfying
\begin{enumerate}
\renewcommand{\labelenumi}{(\alph{enumi})}
   \item $G_s\in\bbD^{1,2}$ for a.e. $s\in[0,T]$,
   \item $\bbE\l[\int_{[0,T]}|G_s|^2ds\r]<\infty$,
   \item $\bbE\l[\int_{[0,T]\times[0,\infty)}\int_0^T|D_{t,z}G_s|^2dsq(dt,dz)\r]<\infty$.
\end{enumerate}
\item
$\bbL_1^{1,2}$ is defined as the space of $G:[0,T]\times(0,\infty)\times\Omega \to\bbR$ such that
\begin{enumerate}
\renewcommand{\labelenumi}{(\alph{enumi})}
\setcounter{enumi}{3}
   \item $G_{s,x}\in\bbD^{1,2}$ for $q$-a.e. $(s,x)\in[0,T]\times(0,\infty)$,
   \item $\bbE\l[\int_{[0,T]\times(0,\infty)}|G_{s,x}|^2\nu(dx)ds\r]<\infty$,
   \item $\bbE\l[\int_{[0,T]\times[0,\infty)}\int_{[0,T]\times(0,\infty)}|D_{t,z}G_{s,x}|^2\nu(dx)dsq(dt,dz)\r]<\infty$.
\end{enumerate}
\item
$\widetilde{\bbL}_1^{1,2}$ is defined as the space of $G\in\bbL_1^{1,2}$ such that
\begin{enumerate}
\renewcommand{\labelenumi}{(\alph{enumi})}
\setcounter{enumi}{6}
   \item $\bbE\l[\l(\int_{[0,T]\times(0,\infty)}|G_{s,x}|\nu(dx)ds\r)^2\r]<\infty$,
   \item $\bbE\l[\int_{[0,T]\times[0,\infty)}\l(\int_{[0,T]\times(0,\infty)}|D_{t,z}G_{s,x}|\nu(dx)ds\r)^2q(dt,dz)\r]<\infty$.
\end{enumerate}
\end{itemize}

\noindent
{\bf Condition C1:}
First, we see $u\in\bbL^{1,2}_0$. To this end, we check items (a)--(c) in the definition of $\bbL^{1,2}_0$.
Lemmas \ref{lem-ut-2} and \ref{lem-ut-1} ensure items (a) and (b), respectively. To see item (c), Lemma \ref{lem-ut-2} implies
\begin{align*}
&\bbE\l[\int_{[0,T]\times[0,\infty)}\int_0^T|D_{t,z}u_s|^2dsq(dt,dz)\r]
\leq \int_{[0,T]\times[0,\infty)}(T-t)\frac{C_u^2}{z}z^2\nu(dz)dt<\infty,
\end{align*}
from which $u\in\bbL_0^{1,2}$ follows.

Next, we show $2u_sD_{t,z}u_s+z(D_{t,z}u_s)^2\in L^2(q\times\bbP)$ as
\begin{align}
\label{eq-cond1-1}
\lefteqn{\bbE\l[\int_{[0,T]\times[0,\infty)}(2u_sD_{t,z}u_s+z(D_{t,z}u_s)^2)^2q(dt,dz)\r]} \nonumber \\
&\leq 2C_u^4\int_{[0,T]\times[0,\infty)}\l(\frac{4}{z}+1\r)z^2\nu(dz)dt
      <\infty
\end{align}
by Lemmas \ref{lem-ut-1} and \ref{lem-ut-2}.

Finally, we prove $u^2\in\bbL_0^{1,2}$. Item (b) holds by Lemma \ref{lem-ut-1}.
As $u_s\in\bbD^{1,2}$ and $u_s^2\in L^2(\bbP)$, Propositions 5.1 and 5.4 of \cite{S07}, together with (\ref{eq-cond1-1}), imply item (a) and $D_{t,z}u_s^2=2u_sD_{t,z}u_s+z(D_{t,z}u_s)^2$.
Moreover, a similar calculation with (\ref{eq-cond1-1}) gives
item (c) as follows:
\begin{align*}
\lefteqn{\bbE\l[\int_{[0,T]\times[0,\infty)}\int_0^T(D_{t,z}u_s^2)^2dsq(dt,dz)\r]} \\
&= \bbE\l[\int_{[0,T]\times[0,\infty)}\int_0^T(2u_sD_{t,z}u_s+z(D_{t,z}u_s)^2)^2dsq(dt,dz)\r]
   <\infty.
\end{align*}
\fin

\noindent
{\bf Condition C2:}
We first demonstrate $\log(1-\theta)\in\bbL_1^{1,2}$. Items (d) and (e) in the definition of $\bbL_1^{1,2}$ are given by Lemmas \ref{lem-ut-4} and \ref{lem-ut-1}, respectively.
As for item (f), Lemmas \ref{lem-ut-3} and \ref{lem-ut-4} imply
\[
|D_{t,z}\log(1-\theta_{s,x})|^2\leq \frac{(C_\theta^\prime)^2}{z}e^{-2\rho x}(1-e^{\rho x})^2.
\]
Because $\int_0^\infty e^{-2\rho x}(1-e^{\rho x})^2\nu(dx)\leq\int_0^1e^{-2\rho} \rho^2x^2\nu(dx)+\int_1^\infty e^{-2\rho x}\nu(dx)<\infty$ by Assumption \ref{ass1}, item (f) follows.

Next, we show $\theta+\log (1-\theta)\in\widetilde{\bbL}_1^{1,2}$.
Note that we can demonstrate $\theta\in\bbL_1^{1,2}$ in the same manner as in the proof of condition C1.
Hence, we have only to see items (g) and (h) in the definition of $\widetilde{\bbL}_1^{1,2}$.
Because $|\theta_{s,x}+\log (1-\theta_{s,x})|\leq 2C_\theta|\rho|x$, item (g) follows.
Next, Lemmas \ref{lem-ut-4} and \ref{lem-ut-3}, and Assumption \ref{ass1} imply
\begin{align*}
&\int_{[0,T]\times(0,\infty)}|D_{t,z}(\theta_{s,x}+\log(1-\theta_{s,x}))|\nu(dx)ds \\
&\leq \int_{[0,T]\times(0,\infty)}|D_{t,z}\theta_{s,x}|(1+e^{-\rho x})\nu(dx)ds \\
&\leq \int_{[0,T]\times(0,\infty)}\frac{C_\theta^\prime}{\sqrt{z}}(1-e^{\rho x})(1+e^{-\rho x})\nu(dx)ds
      \leq \frac{CT}{\sqrt{z}}
\end{align*}
for some $C>0$, from which item (h) follows.
\fin

\noindent
{\bf Condition C3:}
This is given by Lemma \ref{lem-ut-1}.
\fin

\noindent
{\bf Condition C4:}
Proposition \ref{prop-logZT} implies that $\log Z_T\in\bbD^{1,2}$, and $D_{t,0}\log Z_T=u_t$,
from which $\bbE\l[\int_0^T(Z_TD_{t,0}\log Z_T)^2dt\r] <\infty$ follows by Lemma \ref{lem-ut-1} and Proposition \ref{prop-MMM}.
Next, let $\Psi_{t,z}$ be the increment quoting operator defined in \cite{S07}.
That is, for any random variable $F$, $\omega_W\in\Omega_W$ and $\omega_J=((t_1,z_1), \dots,(t_n,z_n))\in\Omega_J$, we define
\[
\Psi_{t,z}F(\omega_W,\omega_J):=\frac{F(\omega_W,\omega_J^{t,z})-F(\omega_W,\omega_J)}{z},
\]
where $\omega_J^{t,z}:=((t,z),(t_1,z_1),\dots,(t_n,z_n))$. As $Z_T\in\bbD^{1,2}$ by Section \ref{sect-ZT}, Proposition 5.4 of \cite{S07} yields that, for $z>0$,
\begin{align}
D_{t,z}Z_T
&= \Psi_{t,z}Z_T=\Psi_{t,z}\exp\{\log Z_T\} \nonumber \\
&= \frac{\exp\{\log Z_T(\omega_W,\omega_J^{t,z})\}-\exp\{\log Z_T(\omega_W,\omega_J)\}}{z} \nonumber \\
&= \frac{\exp\{\log Z_T+z\frac{\log Z_T(\omega_W,\omega_J^{t,z})-\log Z_T(\omega_W,\omega_J)}{z}\}-\exp\{\log Z_T\}}{z} \nonumber \\
&= \frac{\exp\{\log Z_T+z\Psi_{t,z}\log Z_T\}-\exp\{\log Z_T\}}{z} \nonumber \\
&= \frac{\exp\{\log Z_T+zD_{t,z}\log Z_T\}-\exp\{\log Z_T\}}{z} \nonumber \\
&= Z_T\frac{\exp(zD_{t,z}\log Z_T)-1}{z}.
\label{eq-cond4}
\end{align}
As a result, condition C4 follows.
\fin

\noindent
{\bf Condition C5:}
Noting that $|F+zD_{t,z}F|\leq K$ by Theorem \ref{prop-put-deriv}, we have $FD_{t,z}Z_T+zD_{t,z}F\cdot D_{t,z} Z_T\in L^2(q\times\bbP)$, as $Z_T\in\bbD^{1,2}$.
Therefore, it suffices to show $Z_TD_{t,z}F\in L^2(q\times\bbP)$.
To this end, we prove that $\bbE\l[\int_0^T(Z_TD_{t,0}F)^2dt\r]<\infty$ firstly.
Because $D_{t,0}F = -{\bf 1}_{\{S_T<K\}} S_TD_{t,0}L_T = -{\bf 1}_{\{S_T<K\}}S_T\sigma_t$ by Propositions \ref{prop-put-deriv} and \ref{prop-LT-1},
we have $\bbE\l[\int_0^T(Z_TD_{t,0}F)^2dt\r]\leq\bbE\l[Z_T^2K^2\int_0^T\sigma_t^2dt\r]$.
Hence, we have only to show $\bbE[Z_T^2J_T]<\infty$ from the view of (\ref{eq-sigma-2}).
Now, as seen in the proof of Proposition \ref{prop-MMM}, $Y$ defined in (\ref{eq-MMM-2}) is a positive martingale.
Therefore, we can define a probability measure $\bbP_Y$ as $d\bbP_Y=Y_Td\bbP$, and we have
\[
\bbE[Y_TJ_T]=\bbE_{\bbP_Y}[J_T]=\bbE_{\bbP_Y}\l[\int_0^T\int_0^\infty(1-\delta_{s,x})x\nu(dx)ds\r]<\infty,
\]
as $(1-\delta_{s,x})x=(1-\theta_{s,x})^2x\leq(1+C_\theta)^2x$.
Hence, (\ref{eq-MMM-3}) implies that $\bbE[Z_T^2J_T]<\infty$.

Next, we show
$\bbE\l[\int_0^T\int_0^\infty(Z_TD_{t,z}F)^2z^2\nu(dz)dt\r]<\infty$.
Note that
\begin{align*}
\bbE\l[\int_0^T\int_1^\infty(Z_TD_{t,z}F)^2z^2\nu(dz)dt\r]
&\leq \bbE\l[\int_0^T\int_1^\infty\l(Z_T\frac{K}{z}\r)^2z^2\nu(dz)dt\r] \\
&\leq K^2\bbE\l[Z_T^2\int_0^T\int_1^\infty\nu(dz)dt\r]<\infty.
\end{align*}
We have therefore only to show
$\bbE\l[\int_0^T\int_0^1Z_T^2|D_{t,z}F|^2z^2\nu(dz)dt\r]<\infty$.
If we have
\begin{equation}
\label{eq-cond5-2}
|D_{t,z}F|\leq K|D_{t,z}L_T|,
\end{equation}
and there is a $C>0$ such that
\begin{equation}
\label{eq-cond5-3}
\bbE\l[Z_T^2|D_{t,z}L_T|^2\r]<\frac{C}{z}
\end{equation}
for any $z\in(0,1)$, then we obtain
\begin{align*}
\bbE\l[\int_0^T\int_0^1Z_T^2|D_{t,z}F|^2z^2\nu(dz)dt\r]
&\leq K^2\int_0^T\int_0^1\bbE\l[Z_T^2|D_{t,z}L_T|^2\r]z^2\nu(dz)dt \\
&\leq K^2C\int_0^T\int_0^1z\nu(dz)dt<\infty.
\end{align*}

All that remains to show is (\ref{eq-cond5-2}) and (\ref{eq-cond5-3}). (\ref{eq-cond5-2}) follows from
\begin{align*}
|D_{t,z}F|
&=    \frac{|(K-f_K(L_T+zD_{t,z}L_T))^+-(K-f_K(L_T))^+|}{|z|} \\
&\leq \frac{|f_K(L_T+zD_{t,z}L_T))-f_K(L_T)|}{|z|}
      \leq \frac{K|zD_{t,z}L_T|}{|z|}=K|D_{t,z}L_T|.
\end{align*}
Next, to prove (\ref{eq-cond5-3}), it suffices to show that $\bbE_{\bbP^Y}[|D_{t,z}L_T|^2]<Cz^{-1}$ for some $C>0$.
The process $W^Y$ defined as $dW^Y_s:=dW_s+2u_sds$ is a Brownian motion under $\bbP^Y$.
Noting that $\sqrt{\sigma_s^2+ze^{-\lambda(s-t)}}-\sigma_s\leq\sqrt{z}$ for $s\in[t,T]$, we have
\[
|D_{t,z}L_T|\leq C_1+\l|\int_t^T\frac{\sqrt{\sigma_s^2+ze^{-\lambda(s-t)}}-\sigma_s}{z}dW^Y_s\r|+\frac{2C_u(T-t)}{\sqrt{z}}
\]
for some $C_1>0$ by Proposition \ref{prop-LT-1}. Hence, we have
\[
\bbE_{\bbP^Y}[|D_{t,z}L_T|^2]
\leq 3C_1^2+3\bbE_{\bbP^Y}\l[\int_t^T\frac{1}{z}ds\r]+\frac{12C_u^2(T-t)^2}{z}
\leq \frac{C}{z}
\]
for some $C>0$ as $0<z<1$.
\fin

\noindent
{\bf Condition C6:}
To demonstrate $FH^*_{t,z}\in L^1(\bbP^*)$ for $q$-a.e. $(t,z)\in[0,T]\times(0,\infty)$, it suffices to show $\bbE[Z_TH^*_{t,z}]<\infty$, since $F$ is bounded. Now, we have
\[
Z_TH^*_{t,z}
=    Z_T\frac{e^{zD_{t,z}\log Z_T}}{1-\theta_{t,z}}
=    \frac{zD_{t,z}Z_T+Z_T}{1-\theta_{t,z}}
\leq \hat{C}_\theta\{zD_{t,z}Z_T+Z_T\}
\]
by (\ref{eq-cond4}) and item 5 of Lemma \ref{lem-ut-1}. As $Z_T\in\bbD^{1,2}$ by Section \ref{sect-ZT}, we have $D_{t,z}Z_T\in L^1(\bbP)$ for $q$-a.e. $(t,z)\in [0,T]\times(0,\infty)$.
Hence, $\bbE[Z_TH^*_{t,z}]<\infty$. Moreover, because $D_{t,z}F\leq\frac{K}{z}$, we have $H^*_{t,z}D_{t,z}F\in L^1(\bbP^*)$ for $q$-a.e. $(t,z)$.
\fin

\vspace{5mm}

\noindent
{\bf Condition AS3 in Theorem \ref{thm-AS37}:}
As the last part of the proof of Theorem \ref{thm-main}, we make sure of Condition AS3, which is given as follows:
\begin{equation}
\label{eq-thm-AS31}
\bbE\l[\int_0^T\l\{(h^0_t)^2+\int_0^\infty(h^1_{t,z})^2\nu(dz)\r\}dt\r]<\infty,
\end{equation}
where $h^1_{t,z}:=\bbE_{\tP}[F(H^*_{t,z}-1)+zH^*_{t,z}D_{t,z}F|\calF_{t-}]$, and
\begin{align*}
h^0_t
&:= \bbE_{\tP}\l[D_{t,0}F-F\l[\int_0^TD_{t,0}u_sdW^{\tP}_s+\int_0^T\int_0^\infty\frac{D_{t,0}\theta_{s,x}}{1-\theta_{s,x}}\tN^{\tP}(ds,dx)\r]\Big|\calF_{t-}\r] \\
&=  -\bbE_{\tP}\l[{\bf 1}_{\{S_T<K\}}S_T\sigma_t\Big|\calF_{t-}\r].
\end{align*}
Here $dW^{\tP}_t:=dW_t+u_tdt$ and $\tN^{\tP}(dt,dz):=\tN(dt,dz)+\theta_{t,z}\nu(dz)dt$ are a Brownian motion and the compensated Poisson random measure of $N$ under $\tP$, respectively.

First, we have $\bbE\l[\int_0^T(h^0_t)^2dt\r]\leq K^2\bbE \l[\int_0^T\sigma_t^2dt\r]<\infty$ by (\ref{eq-sigma-2}).
Next, we show
$\bbE\l[\int_0^T\int_0^\infty(h^1_{t,z})^2\nu(dz)dt\r]<\infty$.
Noting that $h^1_{t,z}=\bbE_{\tP}[(F+zD_{t,z}F)H^*_{t,z}-F|\calF_{t-}]$, we have
\[
h^1_{t,z}\leq \bbE_{\tP}[(F+zD_{t,z}F)H^*_{t,z}|\calF_{t-}]
         \leq K\bbE_{\tP}[H^*_{t,z}|\calF_{t-}]=K,
\]
as $F$ and $H^*_{t,z}$ are nonnegative, $0\leq F+zD_{t,z}F\leq K$ by Proposition \ref{prop-put-deriv}, and $\bbE_{\tP}[H^*_{t,z}|\calF_{t-}]=1$ by (\ref{eq-thm-main-rem-3}).
In addition, the following holds:
\[
h^1_{t,z}\geq-\bbE_{\tP}[F|\calF_{t-}]\geq-K.
\]
As a result, $h^1_{t,z}$ is bounded.
Hence, we obtain $\bbE\l[\int_0^T\int_1^\infty(h^1_{t,z})^2\nu(dz)dt\r]<\infty$.

Next, we show $\bbE\l[\int_0^T\int_0^1(h^1_{t,z})^2\nu(dz)dt\r]<\infty$.
To this end, we rewrite $h^1_{t,z}$ as
\[
h^1_{t,z}=\bbE_{\tP}[(F+zD_{t,z}F)(H^*_{t,z}-1)+zD_{t,z}F|\calF_{t-}].
\]
Because $|zD_{t,z}F|\leq K$, we have $(\bbE_{\tP}[zD_{t,z}F|\calF_{t-}])^2\leq K^2$.
Thus, it suffices to prove
\begin{equation}
\label{eq-main-1}
\bbE\l[\int_0^T\int_0^1\{\bbE_{\tP}[(F+zD_{t,z}F)(H^*_{t,z}-1)|\calF_{t-}]\}^2\nu(dz)dt\r]<\infty.
\end{equation}
(\ref{eq-thm-main-rem-3}) implies
\begin{align}
\lefteqn{\l\{\bbE_{\tP}\l[(F+zD_{t,z}F)(H^*_{t,z}-1)|\calF_{t-}\r]\r\}^2}
\nonumber \\
&\leq K^2\bbE_{\tP}\l[(H^*_{t,z}-1)^2|\calF_{t-}\r] \nonumber \\
&\leq K^2\l\{\bbE_{\tP}\l[(H^*_{t,z})^2|\calF_{t-}\r]-2\bbE_{\tP}[H^*_{t,z}|\calF_{t-}]+1\r\} \nonumber \\
&=   K^2\l\{\bbE_{\tP}\l[(H^*_{t,z})^2|\calF_{t-}\r]-1\r\}.
\label{eq-main-1-2}
\end{align}
Next, we calculate $(H^*_{t,z})^2$. By the definition of $H^*_{t,z}$ in Theorem \ref{thm-main}, and Proposition \ref{prop-logZT}, we have
\begin{align}
(H^*_{t,z})^2
&=  \exp\bigg\{-2z\int_0^TD_{t,z}u_sdW_s-2z\int_0^Tu_sD_{t,z}u_sds-z^2\int_0^T(D_{t,z}u_s)^2ds \nonumber \\
&   \hspace{5mm}+2z\int_0^T\int_0^\infty D_{t,z}\log(1-\theta_{s,x})\tN(ds,dx) \nonumber \\
&   \hspace{5mm}+2z\int_0^T\int_0^\infty[D_{t,z}\log(1-\theta_{s,x})+D_{t,z}\theta_{s,x}]\nu(dx)ds\bigg\} \nonumber \\
&=  \exp\bigg\{-2z\int_0^TD_{t,z}u_sdW_s-2z\int_0^Tu_sD_{t,z}u_sds-\frac{1}{2}\int_0^T(2zD_{t,z}u_s)^2ds \nonumber \\
&   \hspace{5mm}+\int_0^T(zD_{t,z}u_s)^2ds+\int_0^T\int_0^\infty \log(1-\gamma_{t,z,s,x})\tN(ds,dx) \nonumber \\
&   \hspace{5mm}+\int_0^T\int_0^\infty[\log(1-\gamma_{t,z,s,x})+\gamma_{t,z,s,x}]\nu(dx)ds \nonumber \\
&   \hspace{5mm}-\int_0^T\int_0^\infty\gamma_{t,z,s,x}\theta_{s,x}\nu(dx)ds+\int_0^T\int_0^\infty\frac{(zD_{t,z}\theta_{s,x})^2}{1-\theta_{s,x}}\nu(dx)ds\bigg\},
\label{eq-main-2}
\end{align}
where $\gamma_{t,z,s,x}:=2\frac{zD_{t,z}\theta_{s,x}}{1-\theta_{s,x}}-\l(\frac{zD_{t,z}\theta_{s,x}}{1-\theta_{s,x}}\r)^2$.
We remark that Lemma \ref{lem-ut-4} implies that
\begin{align*}
zD_{t,z}\log(1-\theta_{s,x})
&= \log(1-\theta_{s,x}-zD_{t,z}\theta_{s,x})-\log(1-\theta_{s,x}) \\
&= \log\l(1-\frac{zD_{t,z}\theta_{s,x}}{1-\theta_{s,x}}\r),
\end{align*}
that is, $2zD_{t,z}\log(1-\theta_{s,x})=\log(1-\gamma_{t,z,s,x})$.
Now, we have $(zD_{t,z}u_s)^2\leq zC_u^2$ by Lemma \ref{lem-ut-2}, and
\[
\int_0^\infty\frac{(zD_{t,z}\theta_{s,x})^2}{1-\theta_{s,x}}\nu(dx)\leq z(C_\theta^\prime)^2\hat{C}_\theta C_\rho
\]
by Lemmas \ref{lem-ut-1} and \ref{lem-ut-3}. Therefore, we have
\begin{align}
\lefteqn{\mbox{R.H.S. of (\ref{eq-main-2})}} \nonumber \\
&\leq \exp\bigg\{-2z\int_0^TD_{t,z}u_sdW_s-2z\int_0^Tu_sD_{t,z}u_sds-\frac{1}{2}\int_0^T(2zD_{t,z}u_s)^2ds \nonumber \\
&     \hspace{5mm}+\int_0^T\int_0^\infty \log(1-\gamma_{t,z,s,x})\tN(ds,dx)+\int_0^T\int_0^\infty[\log(1-\gamma_{t,z,s,x})+\gamma_{t,z,s,x}]\nu(dx)ds \nonumber \\
&     \hspace{5mm}-\int_0^T\int_0^\infty\gamma_{t,z,s,x}\theta_{s,x}\nu(dx)ds+Cz\bigg\}
\label{eq-main-3}
\end{align}
for some $C>0$.
Hence, Lemma \ref{lem-main-1} implies that
\begin{equation}
\label{eq-main-4}
\bbE_{\tP}\l[(H^*_{t,z})^2|\calF_{t-}\r]
\leq \bbE_{\tP}\l[X^{t,z}_T|\calF_{t-}\r]e^{Cz}
=    X^{t,z}_{t-}e^{Cz}=e^{Cz}.
\end{equation}
Consequently, we have
\[
\mbox{R.H.S. of (\ref{eq-main-1-2})}\leq K^2\l(e^{Cz}-1\r)\leq K^2z\l(e^C-1\r)
\]
for any $z\in(0,1)$.
Hence, (\ref{eq-main-1}) follows, from which we obtain (\ref{eq-thm-AS31}). This completes the proof of Theorem \ref{thm-main}.
\fin

To see (\ref{eq-main-4}), we show the following lemma.

\begin{lem}
\label{lem-main-1}
Given $(t,z)\in[0,T]\times(0,\infty)$, we consider the following SDE:
\begin{align}
\label{eq-lem-main-1-1}
dX^{t,z}_s
&= -X^{t,z}_{s-}\bigg\{2zD_{t,z}u_sdW_s+2zu_sD_{t,z}u_sds+\int_0^\infty\gamma_{t,z,s,x}\tN(ds,dx) \nonumber \\
&  \hspace{7mm}+\int_0^\infty\gamma_{t,z,s,x}\theta_{s,x}\nu(dx)ds\bigg\}.
\end{align}
Then, the solution $X^{t,z}$ is a martingale under $\tP$ with $X^{t,z}_s=1$ for any $s\in[0,t)$.
In particular, the right-hand side of (\ref{eq-main-3}) is equal to $X^{t,z}_Te^{Cz}$.
\end{lem}

\proof
First, note that $zD_{t,z}u_s$ and $zu_sD_{t,z}u_s$ are bounded. In addition, we have
\begin{equation}
\label{eq-lem-main-1-2}
\l|\frac{zD_{t,z}\theta_{s,x}}{1-\theta_{s,x}}\r| < 2C_\theta\hat{C}_\theta(1-e^{\rho x}) < 2C_\theta\hat{C}_\theta
\end{equation}
by Lemmas \ref{lem-ut-1} and \ref{lem-ut-3}. Therefore, Lemma \ref{lem-ut-1} yields
\begin{align*}
\int_0^\infty|\gamma_{t,z,s,x}\theta_{s,x}|\nu(dx)
&=    \int_0^\infty\l|\frac{zD_{t,z}\theta_{s,x}}{1-\theta_{s,x}}\l(2-\frac{zD_{t,z}\theta_{s,x}}{1-\theta_{s,x}}\r)\theta_{s,x}\r|\nu(dx) \\
&\leq 2C_\theta\hat{C}_\theta(2+2C_\theta\hat{C}_\theta)\cdot C_\theta|\rho|\int_0^\infty x\nu(dx)
      <\infty.
\end{align*}
Moreover, (\ref{eq-lem-main-1-2}) again implies
\begin{align*}
\int_0^\infty\gamma^2_{t,z,s,x}\nu(dx)
&=    \int_0^\infty\frac{(zD_{t,z}\theta_{s,x})^2}{(1-\theta_{s,x})^2}\l(2-\frac{zD_{t,z}\theta_{s,x}}{1-\theta_{s,x}}\r)^2\nu(dx) \\
&\leq 4C^2_\theta\hat{C}^2_\theta C_\rho(2+2C_\theta\hat{C}_\theta)^2.
\end{align*}
As a result, we can apply Theorem 117 of Situ \cite{Situ} to (\ref{eq-lem-main-1-1});
we then conclude that (\ref{eq-lem-main-1-1}) has a solution $X^{t,z}$ satisfying $\bbE\l[\sup_{t\leq s\leq T}|X^{t,z}_s|^2\r]<\infty$,
which implies $\bbE_{\tP}[\sup_{t\leq s\leq T}|X^{t,z}_s|]<\infty$ by the $L^2(\bbP)$-property of $Z_T$.
Now, $X^{t,z}$ is a local martingale under $\tP$, because we can rewrite (\ref{eq-lem-main-1-1}) as
\[
dX^{t,z}_s= -X^{t,z}_{s-}\l\{2zD_{t,z}u_sdW^{\tP}_s+\int_0^\infty\gamma_{t,z,s,x}\tN^{\tP}(ds,dx)\r\}.
\]
Consequently, Theorem I.51 of Protter \cite{P} implies that $X^{t,z}$ is a $\tP$-martingale satisfying $X^{t,z}_s=1$ for any $s\in[0,t)$.
Moreover, by Example 9.6 of Di Nunno et al. \cite{NOP}, the right-hand side of (\ref{eq-main-3}) is expressed by $X^{t,z}_Te^{Cz}$.
\fin

\setcounter{equation}{0}
\section{Malliavin differentiability of $Z$}
\label{sect-ZT}
This section is devoted to show $Z_t\in\bbD^{1,2}$ for any $t\in[0,T]$. To this end, for $t\in[0,T]$, we define $Z_t^{(0)}:=1$ and
\[
Z_t^{(n+1)}:=1-\int_0^tZ_{s-}^{(n)}u_sdW_s-\int_0^t\int_0^\infty Z_{s-}^{(n)}\theta_{s,x}\tN(ds,dx)
\]
for $n\geq 0$. Furthermore, we denote, for $n\geq 0$,
\[
\phi_n(t):=\bbE\l[\int_{[0,t]\times[0,\infty)}\l(D_{r,z}Z_t^{(n)}\r)^2q(dr,dz)\r].
\]
Note that $\phi_0(t)\equiv0$.

\begin{lem}
\label{lem-ZT}
We have $Z_t^{(n)}\in\bbD^{1,2}$ for every $n\geq 0$ and any $t\in [0,T]$.
Moreover, there exist constants $k_1>0$ and $k_2>0$ such that
\[
\phi_{n+1}(t)\leq k_1+k_2\int_0^t\phi_n(s)ds
\]
for every $n\geq0$ and any $t\in[0,T]$.
\end{lem}

\noindent
Under Lemma \ref{lem-ZT}, we have
\begin{align*}
\phi_{n+1}(t)
&\leq k_1+k_2\int_0^t\phi_n(s)ds
      \leq k_1+k_2\int_0^t\l(k_1+k_2\int_0^s\phi_{n-1}(s_1)ds_1\r)ds \\
&\leq \cdots\leq k_1\sum_{j=0}^n\frac{k_2^jt^j}{j!}<k_1e^{k_2t}.
\end{align*}
for any $t\in[0,T]$. Hence, $\sup_{n\geq1}\phi_n(t)<\infty$ holds.
As $Z^{(n)}_t\to Z_t$ in $L^2(\bbP)$, Lemma 17.1 of \cite{NOP} implies that $Z_t\in\bbD^{1,2}$ for $t\in [0,T]$.
Note that the Malliavin derivative in \cite{NOP} is defined in a different way from ours.
Denoting by $\hat{D}$ the Malliavin derivative operator in \cite{NOP}, we have $\hat{D}_{t,z}F=zD_{t,z}F$ for $z\neq0$ and $F\in\bbD^{1,2}$.

{\it Proof of Lemma \ref{lem-ZT}.}$\quad$$\quad$
We take an integer $n\geq0$ arbitrarily.
Suppose that $Z_t^{(n)}\in\bbD^{1,2}$ and $\int_0^t\phi_n(s)ds<\infty$ for any $t\in[0,T]$.
Lemma \ref{lem-ZT-2} below and Lemma 3.3 of \cite{DI} imply that $Z_t^{(n+1)}\in\bbD^{1,2}$ for any $t\in[0,T]$; and, for any $t\in[r,T]$ and any $z\in(0,\infty)$,
\begin{align}
D_{r,0}Z_t^{(n+1)}
&= -D_{r,0}\int_{[0,T]\times[0,\infty)}Z_{s-}^{(n)}\l\{u_s{\bf 1}_{\{0\}}(x)+\frac{\theta_{s,x}}{x}{\bf 1}_{(0,\infty)}(x)\r\}{\bf 1}_{[0,t]}(s)Q(ds,dx) \nonumber \\
&= -Z_{r-}^{(n)}u_r-\int_r^tD_{r,0}(Z_{s-}^{(n)}u_s)dW_s-\int_r^t\int_0^\infty D_{r,0}\l(Z_{s-}^{(n)}\frac{\theta_{s,x}}{x}\r)x\tN(ds,dx) \nonumber \\
&= -Z_{r-}^{(n)}u_r-\int_r^tu_sD_{r,0}Z_{s-}^{(n)}dW_s-\int_r^t\int_0^\infty\theta_{s,x}D_{r,0}Z_{s-}^{(n)}\tN(ds,dx)
\label{eq-ZT-0}
\end{align}
and
\begin{equation}
\label{eq-ZT-0-z}
D_{r,z}Z_t^{(n+1)}= -Z_{r-}^{(n)}\frac{\theta_{r,z}}{z}-\int_r^tD_{r,z}\l(Z_{s-}^{(n)}u_s\r)dW_s-\int_r^t\int_0^\infty D_{r,z}\l(Z_{s-}^{(n)}\frac{\theta_{s,x}}{x}\r)x\tN(ds,dx).
\end{equation}
Next, we fix $t\in[0,T]$ arbitrarily. We have then
\begin{equation}
\label{eq-ZT-1}
\phi_{n+1}(t)= \bbE\l[\int_0^t\l(D_{r,0}Z_t^{(n+1)}\r)^2dr\r]+\bbE\l[\int_0^t\int_0^\infty\l(D_{r,z}Z_t^{(n+1)}\r)^2z^2\nu(dz)dr\r].
\end{equation}
(\ref{eq-ZT-0}) implies
\begin{align}
\label{eq-ZT-2}
\lefteqn{\mbox{The first term of (\ref{eq-ZT-1})}} \nonumber \\
&\leq 3\bbE\l[\int_0^t\l(Z_{r-}^{(n)}u_r\r)^2dr\r]+3\bbE\l[\int_0^t\l(\int_r^tu_sD_{r,0}Z_{s-}^{(n)}dW_s\r)^2dr\r]\nonumber \\
&     \hspace{5mm}+3\bbE\l[\int_0^t\l(\int_r^t\int_0^\infty\theta_{s,x}D_{r,0}Z_{s-}^{(n)}\tN(ds,dx)\r)^2dr\r].
\end{align}
We evaluate each term on the right-hand side of (\ref{eq-ZT-2}). Lemma \ref{lem-ut-1} implies
\[
\bbE\l[\int_0^t\l(Z_{r-}^{(n)}u_r\r)^2dr\r]
\leq C_u^2\bbE\l[\int_0^t\l(Z_{r-}^{(n)}\r)^2dr\r]
\leq C_u^2T\bbE\l[\sup_{0\leq s\leq T}\l(Z_s^{(n)}\r)^2\r]
\]
and
\[
\bbE\l[\int_0^t\l(\int_r^tu_sD_{r,0}Z_{s-}^{(n)}dW_s\r)^2dr\r]\leq C_u^2\int_0^t\bbE\l[\int_r^t\l(D_{r,0}Z_{s-}^{(n)}\r)^2ds\r]dr.
\]
The same argument implies that
\begin{align*}
\lefteqn{\bbE\l[\int_0^t\l(\int_r^t\int_0^\infty\theta_{s,x}D_{r,0}Z_{s-}^{(n)}\tN(ds,dx)\r)^2dr\r]} \\
&= \int_0^t\bbE\l[\int_r^t\int_0^\infty\l(\theta_{s,x}D_{r,0}Z_{s-}^{(n)}\r)^2\nu(dx)ds\r]dr
   \leq C_\theta^2C_\rho\int_0^t\bbE\l[\int_r^t\l(D_{r,0}Z_{s-}^{(n)}\r)^2ds\r]dr.
\end{align*}
As a result, we obtain
\begin{align}
\label{eq-ZT-3}
\lefteqn{\mbox{The first term of (\ref{eq-ZT-1})}} \nonumber \\
&\leq 3C_u^2T\bbE\l[\sup_{0\leq s\leq T}\l(Z_s^{(n)}\r)^2\r]+3(C_u^2+C_\theta^2C_\rho)\int_0^t\bbE\l[\int_r^t\l(D_{r,0}Z_{s-}^{(n)}\r)^2ds\r]dr.
\end{align}
Next, (\ref{eq-ZT-0-z}) yields
\begin{align}
\label{eq-ZT-4}
\lefteqn{\mbox{The second term of (\ref{eq-ZT-1})}} \nonumber \\
&\leq 3\bbE\l[\int_0^t\int_0^\infty\l(Z_{r-}^{(n)}\frac{\theta_{r,z}}{z}\r)^2z^2\nu(dz)dr\r] \nonumber \\
&     \hspace{5mm}+3\bbE\l[\int_0^t\int_0^\infty\l(\int_r^tD_{r,z}\l(Z_{s-}^{(n)}u_s\r)dW_s\r)^2z^2\nu(dz)dr\r]\nonumber \\
&     \hspace{5mm}+3\bbE\l[\int_0^t\int_0^\infty\l(\int_r^t\int_0^\infty D_{r,z}\l(Z_{s-}^{(n)}\frac{\theta_{s,x}}{x}\r)x\tN(ds,dx)\r)^2z^2\nu(dz)dr\r].
\end{align}
We now calculate each term on the right-hand side of (\ref{eq-ZT-4}).
\begin{equation}
\label{eq-ZT-5}
\mbox{The first term of (\ref{eq-ZT-4})} \leq 3C_\theta^2C_\rho\bbE\l[\int_0^t\l(Z_{r-}^{(n)}\r)^2dr\r] \leq 3C_\theta^2C_\rho T\bbE\l[\sup_{0\leq s\leq T}\l(Z_s^{(n)}\r)^2\r].
\end{equation}
Next, Lemma \ref{lem-ut-2} implies
\begin{align}
\lefteqn{\mbox{The second term of (\ref{eq-ZT-4})}} \nonumber \\
&=    3\int_0^t\int_0^\infty\bbE\l[\int_r^t\l(D_{r,z}\l(Z_{s-}^{(n)}u_s\r)\r)^2ds\r]z^2\nu(dz)dr \nonumber \\
&=    3\int_0^t\int_0^\infty\bbE\l[\int_r^t\l(u_sD_{r,z}Z_{s-}^{(n)}+Z_{s-}^{(n)}D_{r,z}u_s+zD_{r,z}Z_{s-}^{(n)}\cdot D_{r,z}u_s\r)^2ds\r]z^2\nu(dz)dr \nonumber \\
&\leq 9\int_0^t\int_0^\infty\bigg\{C_u^2\bbE\l[\int_r^t\l(D_{r,z}Z_{s-}^{(n)}\r)^2ds\r]+\frac{C_u^2}{z}\bbE\l[\int_r^t\l(Z_{s-}^{(n)}\r)^2ds\r] \nonumber \\
&     \hspace{5mm}+(C^\prime_u)^2\bbE\l[\int_r^t\l(D_{r,z}Z_{s-}^{(n)}\r)^2ds\r]\bigg\}z^2\nu(dz)dr\nonumber \\
&\leq 9C_u^2\int_0^\infty z\nu(dz)T^2\bbE\l[\sup_{0\leq s\leq T}\l(Z_s^{(n)}\r)^2\r] \nonumber \\
&     \hspace{5mm}+9\l(C_u^2+(C^\prime_u)^2\r)\int_0^t\int_0^\infty\bbE\l[\int_r^t\l(D_{r,z}Z_{s-}^{(n)}\r)^2ds\r]z^2\nu(dz)dr.
\label{eq-ZT-6}
\end{align}
Moreover, we evaluate the third term of (\ref{eq-ZT-4}). By Lemma \ref{lem-ut-3}, we obtain
\begin{align}
\lefteqn{\mbox{The third term of (\ref{eq-ZT-4})}} \nonumber \\
&=    3\int_0^t\int_0^\infty\bbE\bigg[\int_r^t\int_0^\infty\bigg\{D_{r,z}Z_{s-}^{(n)}\cdot\frac{\theta_{s,x}}{x}+Z_{s-}^{(n)}D_{r,z}\frac{\theta_{s,x}}{x} \nonumber \\
&     \hspace{5mm}+zD_{r,z}Z_{s-}^{(n)}\cdot D_{r,z}\frac{\theta_{s,x}}{x}\bigg\}^2x^2\nu(dx)ds\bigg]z^2\nu(dz)dr \nonumber \\
&\leq 9\int_0^t\int_0^\infty\bigg\{C_\theta^2C_\rho\bbE\l[\int_r^t\l(D_{r,z}Z_{s-}^{(n)}\r)^2ds\r]+\frac{(C_\theta^\prime)^2C_\rho}{z}\bbE\l[\int_r^t\l(Z_{s-}^{(n)}\r)^2ds\r] \nonumber \\
&     \hspace{5mm}+z^2\frac{4C_\theta^2C_\rho}{z^2}\bbE\l[\int_r^t\l(D_{r,z}Z_{s-}^{(n)}\r)^2ds\r]\bigg\}z^2\nu(dz)dr \nonumber \\
&\leq 9(C_\theta^\prime)^2C_\rho\int_0^\infty z\nu(dz)T^2\bbE\l[\sup_{0\leq s\leq T}\l(Z_s^{(n)}\r)^2\r] \nonumber \\
&     \hspace{5mm}+45C_\theta^2C_\rho\int_0^t\int_0^\infty\bbE\l[\int_r^t\l(D_{r,z}Z_{s-}^{(n)}\r)^2ds\r]z^2\nu(dz)dr.
\label{eq-ZT-7}
\end{align}
Consequently, by (\ref{eq-ZT-1}), (\ref{eq-ZT-3})--(\ref{eq-ZT-7}) and Lemma \ref{lem-ZT-3} below, there are constants $k_1>0$ and $k_2>0$ such that
\begin{align*}
\phi_{n+1}(t)
&\leq k_1\bbE\l[\sup_{0\leq s\leq T}\l(Z_s^{(n)}\r)^2\r]+k_2\int_{[0,t]\times[0,\infty)}\bbE\l[\int_r^t\l(D_{r,z}Z_{s-}^{(n)}\r)^2ds\r]q(dr,dz) \\
&\leq k_1\sup_{n\geq1}\bbE\l[\sup_{0\leq s\leq T}\l(Z_s^{(n)}\r)^2\r]+k_2\int_0^t\bbE\l[\int_{[0,s]\times[0,\infty)}\l(D_{r,z}Z_{s-}^{(n)}\r)^2q(dr,dz)\r]ds \\
&=    k_1\sup_{n\geq1}\bbE\l[\sup_{0\leq s\leq T}\l(Z_s^{(n)}\r)^2\r]+k_2\int_0^t\bbE\l[\int_{[0,s]\times[0,\infty)}\l(D_{r,z}Z_s^{(n)}\r)^2q(dr,dz)\r]ds \\
&\leq k_1+k_2\int_0^t\phi_n(s)ds,
\end{align*}
where $k_1$ and $k_2$ may vary from line to line.
\fin

Now, we prove two lemmas which are used in the proof of Lemma \ref{lem-ZT}.

\begin{lem}
\label{lem-ZT-2}
Fix $n\geq0$ arbitrarily.
Assume that $Z_t^{(n)}\in\bbD^{1,2}$ and $\int_0^t\phi_n(s)ds<\infty$ for any $t\in[0,T]$.
We have $Z_-^{(n)}u \in\bbL_0^{1,2}$ and $Z_-^{(n)}\theta\in \bbL_1^{1,2}$.
\end{lem}

\proof
We show $Z_-^{(n)}u\in\bbL_0^{1,2}$. By $Z_t^{(n)}\in\bbD^{1,2}$, and Lemmas \ref{lem-ut-1} and \ref{lem-ut-2},
we have $Z_{s-}^{(n)}D_{t,z}u_s+u_sD_{t,z} Z_{s-}^{(n)}+zD_{t,z}Z_{s-}^{(n)}\cdot D_{t,z}u_s\in L^2(q\times\bbP)$ for any $s\in[0,T]$.
Hence, item (a) in the definition of $\bbL_0^{1,2}$ is given by Propositions 5.1 and 5.4 of \cite{S07}.
Next, item (b) is satisfied by Lemma \ref{lem-ut-1}.
As for item (c), there exist two constants $C_1>0$ and $C_2>0$ such that $(D_{t,z}(Z^{(n)}_{s-}u_s))^2\leq\frac{C_1}{z}(Z^{(n)}_{s-})^2 +C_2(D_{t,z} Z^{(n)}_{s-})^2$.
In addition, we have
\begin{align*}
\lefteqn{\bbE\l[\int_{[0,T]\times[0,\infty)}\int_0^T\l(D_{t,z}Z^{(n)}_{s-}\r)^2dsq(dt,dz)\r]} \\
&= \int_0^T\bbE\l[\int_{[0,T]\times[0,\infty)}\l(D_{t,z}Z^{(n)}_{s-}\r)^2q(dt,dz)\r]ds = \int_0^T\phi_n(s)ds<\infty.
\end{align*}
As a result, item (c) follows.
This completes the proof of $Z_-^{(n)}u\in \bbL_0^{1,2}$. $Z_-^{(n)}\theta \in\bbL_1^{1,2}$ is shown similarly.
\fin

\begin{lem}
\label{lem-ZT-3}
$\sup_{n\geq1}\bbE\l[\sup_{0\leq s\leq T}\l(Z_s^{(n)}\r)^2\r]<\infty$.
\end{lem}

\proof
First, we can see inductively that $Z^{(n)}$ is a martingale with $Z^{(n)}_T\in L^2(\bbP)$.
Denoting $\zeta_n(t) := \bbE\l[\sup_{0\leq s\leq t}\l(Z_s^{(n)}\r)^2 \r]$ for $t\in[0,T]$ and $n\geq1$, we have
\begin{align*}
\zeta_n(T)
&\leq 4\bbE\l[\l\{1-\int_0^TZ_{s-}^{(n-1)}u_sdW_s-\int_0^T\int_0^\infty Z_{s-}^{(n-1)}\theta_{s,x}\tN(ds,dx)\r\}^2\r] \\
&\leq 4\l\{1+\bbE\l[\int_0^T\l(Z_{s-}^{(n-1)}\r)^2\l\{u_s^2+\int_0^\infty\theta_{s,x}^2\nu(dx)\r\}ds\r]\r\} \\
&\leq 4+4(C_u^2+C_\theta^2C_\rho)\int_0^T\zeta_{n-1}(s)ds \leq 4\exp\{4(C_u^2+C_\theta^2C_\rho)T\}
\end{align*}
by Doob's inequality and Lemma \ref{lem-ut-1}.
\fin

\setcounter{equation}{0}
\section{Numerical experiments}
In this section, we illustrate LRM strategies for the BNS models with numerical experiments.
\cite{AIS} developed a numerical scheme of LRM strategies for exponential L\'evy models using the Carr-Madan approach \cite{CM},
which is a numerical method for option prices based on the fast Fourier transform (FFT).
In the following, we shall compute (\ref{eq-LRM}) numerically for the call options using the method developed in \cite{AIS}.
Moreover, we compare LRM strategies with delta-hedging strategies, which are given as the partial derivative of the option price with respect to the asset price.

We treat the Gamma-OU model in which the L\'evy measure $\nu$ is given as
\[
\nu(dx)=ab\lambda e^{-bx}{\bf 1}_{(0,\infty)}(x)dx,
\]
where $a>0$, $b>0$.
Moreover, we use the parameter set estimated in \cite{Scho} (see Table 1).
To do it, we need to adopt the same setting as \cite{Scho}.
Hence, we need to take into account the interest rate $r>0$ and the continuous dividend rate $q>0$;
that is, the discount factor is given by $r-q$.
Moreover, suppose that the discounted asset price process $e^{-(r-q)t}S_t$ is a martingale.
Hence, $\mu$ appearing in (\ref{eq-S}) is given as
\[
\mu=r-q+\int_0^\infty(1-e^{\rho x})\nu(dx)=r-q-\frac{a\lambda\rho}{b-\rho}.
\]
We consider a call option with strike price $K$. From the view of Theorem \ref{thm-main}, Corollary \ref{cor1} and Proposition \ref{prop-put-deriv}, we have
\begin{align}
\xi_t^{(S_T-K)^+}
&= \frac{e^{-(r-q)(T-t)}}{S_{t-}(\sigma_t^2+C_\rho)}\bigg\{\sigma_t^2\bbE[S_T{\bf 1}_{\{S_T\geq K\}}|\calF_{t-}] \nonumber \\
&  \hspace{5mm}+\int_0^\infty\bbE\l[\l(S_Te^{zD_{t,z}L_T}-K\r)^+-(S_T-K)^+|\calF_{t-}\r](e^{\rho z}-1)\nu(dz)\bigg\},
\label{eq-LRM}
\end{align}
as $H^*_{t,z}=1$ and $Z_T=1$. Therefore, denoting
\[
I_1:= e^{-(r-q)(T-t)}\bbE[S_T{\bf 1}_{\{S_T\geq K\}}|S_t, \sigma^2_t]
\]
and
\[
I_2:= e^{-(r-q)(T-t)}\int_0^\infty\bbE\l[\l(S_Te^{zD_{t,z}L_T}-K\r)^+-(S_T-K)^+|S_t, \sigma^2_t\r](e^{\rho z}-1)\nu(dz),
\]
we can rewrite (\ref{eq-LRM}) as
\[
\xi_t^{(S_T-K)^+}= \frac{\sigma^2_tI_1+I_2}{S_t(\sigma_t^2+C_\rho)}.
\]
We shall next develop numerical schemes for $I_1$ and $I_2$ separately.

Denoting by $\phi$ the characteristic function of $L_T$ given $S_t$ and $\sigma^2_t$, we have
\begin{eqnarray}
\phi(\vt)
&:=& \bbE[\exp\{i\vt L_T\}|S_t,\sigma^2_t] \nonumber \\
&=&  \exp\Bigg\{i\vt\l(L_t+\mu(T-t)\r)-(\vt^2+i\vt)\frac{\calB(T-t)}{2}\sigma^2_t \nonumber \\
&&   \hspace{5mm}+\frac{a}{b-f_2}\l[b\log\l(\frac{b-f_1}{b-i\vt\rho}\r)+f_2\lambda(T-t)\r]\Bigg\}
\label{eq-phi}
\end{eqnarray}
for $\vt\in\bbC$ from Section 7.1.1 in \cite{Scho}, where
\[
f_1:=i\vt\rho-\frac{1}{2}(\vt^2+i\vt)\lambda\calB(T-t) \ \mbox{ and } \
f_2:=i\vt\rho-\frac{1}{2}(\vt^2+i\vt).
\]
Recall that $\calB(t)=\frac{1-e^{-\lambda t}}{\lambda}$ for $t\in[0,T]$. As for $I_1$, Proposition 2.1 of \cite{AIS} implies
\begin{equation}
\label{eq-I1}
I_1=\frac{e^{-(r-q)(T-t)}}{\pi}\int_0^\infty K^{-i\zeta+1}\frac{\phi(\zeta)}{i\zeta-1}dv,
\end{equation}
where $\zeta:=v-i\alpha$, and $\alpha$ is a real number satisfying
\begin{equation}
\label{eq-alpha-cond}
\sup_{t\leq s<T}\l\{\frac{1}{2}-\frac{\rho}{\calB(T-s)}-\sqrt{D_s}\r\}<\alpha<\inf_{t\leq s<T}\l\{\frac{1}{2}-\frac{\rho}{\calB(T-s)}+\sqrt{D_s}\r\}
\end{equation}
by Theorem 2.2 of \cite{NV}. Here,
\[
D_s:=\l(-\frac{1}{2}+\frac{\rho}{\calB(T-s)}\r)^2+\frac{2\hat{\vt}}{\calB(T-s)}
\]
and
\[
\hat{\vt}:=\sup\l\{\vt\in\bbR\Big|\int_0^\infty(e^{\vt x}-1)\nu(dx)<\infty\r\},
\]
which is ensured to be positive by Assumption \ref{ass1}.
Note that the right-hand side of (\ref{eq-I1}) is independent of the choice of $\alpha$.
As a result, since the integrand of (\ref{eq-I1}) is given by the product of $K^{-i\zeta+1}$ and a function of $\zeta$, we can compute $I_1$ through the FFT.

Next, we calculate $I_2$.
First, Proposition \ref{prop-LT-1} implies
\begin{eqnarray*}
\lefteqn{\frac{S_T}{S_t}\exp\{zD_{t,z}L_T\}} \\
&=& \exp\Bigg\{\mu(T-t)-\frac{1}{2}\int_t^T\sigma^2_sds+\int_t^T\sigma_sdW_s+\rho\int_t^TdJ_s \\
&&  \hspace{5mm}-\frac{z}{2}\calB(T-t)+\int_t^T\l(\sqrt{\sigma^2_s+ze^{-\lambda(s-t)}}-\sigma_s\r)dW_s+\rho z\Bigg\} \\
&=& \exp\Bigg\{\mu(T-t)-\frac{1}{2}\int_t^T\l(\sigma^2_s+ze^{-\lambda(s-t)}\r)ds \\
&&  \hspace{5mm}+\int_t^T\sqrt{\sigma^2_s+ze^{-\lambda(s-t)}}dW_s+\rho\int_t^TdJ_s+\rho z\Bigg\} \\
&=& \exp\Bigg\{\mu(T-t)-\frac{1}{2}\int_t^T\sigma^2_{s,z}ds+\int_t^T\sigma_{s,z}dW_s+\rho\int_t^TdJ_s+\rho z\Bigg\}
\end{eqnarray*}
for $t\in[0,T]$ and $z\in(0,\infty)$, where $\calB(T-t)=\int_t^Te^{-\lambda(s-t)}ds$ for $t\in[0,T]$,
and $\sigma^2_{s,z}:=\sigma^2_s+ze^{-\lambda(s-t)}$ for $(s,z)\in[t,T]\times(0,\infty)$.
Denoting
\[
L^{(z)}_s:=\int_t^s\l(\mu-\frac{1}{2}\sigma_{u,z}^2\r)du+\int_t^s\sigma_{u,z}dW_u+\rho\int_t^sdJ_u
\]
for $(s,z)\in[t,T]\times(0,\infty)$, we have
\[
S_T\exp\{zD_{t,z}L_T\}=S_t\exp\{L^{(z)}_T+\rho z\}.
\]
In addition, as the process $(\sigma^2_{s,z})_{t\leq s\leq T}$ is a solution to the SDE (\ref{SDE-sigma}) with $\sigma_{t,z}^2=\sigma^2_t+z$,
(\ref{eq-phi}) implies that the characteristic function of $\log(S_t)+L^{(z)}_T$ given $S_t$ and $\sigma^2$ is described as follows:
\begin{align*}
\phi^{(z)}(\vt)
&:= \bbE\l[\exp\{i\vt L^{(z)}_T\}|S_t,\sigma^2_t\r]S^{i\vt}_t
    =\bbE\l[\exp\{i\vt L_T\}|S_t,\sigma^2_t+z\r] \\
&=  \phi(\vt)\exp\l\{-(\vt^2+i\vt)\frac{\calB(T-t)}{2}z\r\}.
\end{align*}
Proposition 2.3 of \cite{AIS} implies
\begin{eqnarray*}
\lefteqn{e^{(r-q)(T-t)}I_2} \\
&=& \int_0^\infty\bbE\l[\l(S_t\exp\l\{L^{(z)}_T+\rho z\r\}-K\r)^+-(S_T-K)^+\Big|S_t, \sigma^2_t\r](e^{\rho z}-1)\nu(dz) \\
&=& \int_0^\infty\Bigg\{\frac{e^{\rho z}}{\pi}\int_0^\infty(Ke^{-\rho z})^{-i\zeta+1}\frac{\phi^{(z)}(\zeta)}{(i\zeta-1)i\zeta}dv
    -\frac{1}{\pi}\int_0^\infty\frac{K^{-i\zeta+1}\phi(\zeta)}{(i\zeta-1)i\zeta}dv\Bigg\}(e^{\rho z}-1)\nu(dz) \\
&=& \int_0^\infty\frac{1}{\pi}\int_0^\infty\frac{K^{-i\zeta+1}\phi(\zeta)}{(i\zeta-1)i\zeta}
    \Bigg\{e^{i\rho z\zeta}\exp\l\{-(\zeta^2+i\zeta)\frac{\calB(T-t)}{2}z\r\}-1\Bigg\}dv(e^{\rho z}-1)\nu(dz) \\
&=& \int_0^\infty\frac{1}{\pi}\frac{K^{-i\zeta+1}\phi(\zeta)}{(i\zeta-1)i\zeta}\int_0^\infty(e^{\eta z}-1)(e^{\rho z}-1)\nu(dz)dv,
\end{eqnarray*}
where $\eta:=i\rho\zeta-(\zeta^2+i\zeta)\frac{\calB(T-t)}{2}$, which is a function of $\zeta$.
Note that, as in the proof of Theorem 2.2 of \cite{NV}, $\Re(\eta)\leq0$ when $0<\alpha<1-\frac{2\rho}{\calB(T)}$, which is a subinterval of (\ref{eq-alpha-cond}) for any $t\in[0,T]$.
Therefore, taking such an $\alpha$, we have
\[
\int_0^\infty(e^{\eta z}-1)(e^{\rho z}-1)\nu(dz)=ab\lambda\l[\frac{1}{b-\eta-\rho}-\frac{1}{b-\eta}-\frac{1}{b-\rho}+\frac{1}{b}\r],
\]
from which we can compute $I_2$ using the FFT.

Next, we discuss delta-hedging strategy $\Delta_t^{(S_T-K)^+}$ for a call option with strike price $K$, which is given as the partial derivative of the option price with respect to $S_t$,
that is,
\[
\Delta_t^{(S_T-K)^+}:=e^{-(r-q)(T-t)}\frac{\partial}{\partial S_t}\bbE[(S_T-K)^+|S_t,\sigma^2_t].
\]
Noting that
\[
\bbE[(S_T-K)^+|S_t,\sigma^2_t]=\frac{1}{\pi}\int_0^\infty K^{-i\zeta+1}\frac{\phi(\zeta)}{(i\zeta-1)i\zeta}dv,
\]
we have
\begin{align*}
\Delta_t^{(S_T-K)^+}
&= \frac{e^{-(r-q)(T-t)}}{\pi}\int_0^\infty\frac{K^{-i\zeta+1}}{(i\zeta-1)i\zeta}\frac{\partial \phi(\zeta)}{\partial S_t}dv \\
&= \frac{e^{-(r-q)(T-t)}}{\pi}\int_0^\infty K^{-i\zeta+1}\frac{\phi(\zeta)S_t^{-1}}{i\zeta-1}dv
   =\frac{I_1}{S_t}.
\end{align*}
Hence, the delta-hedging strategy is given from $I_1$.

We show numerical results on LRM strategies $\xi_t^{(S_T-K)^+}$ and delta-hedging strategies $\Delta_t^{(S_T-K)^+}$ using the parameter set estimated in \cite{Scho}.
We fix $T=1$, $r=0.019$ and $q=0.012$.
The asset price and the squared volatility at time $t$ are fixed to $S_t=1124.47$ and $\sigma_t^2=0.0145$, respectively.
Recall Table 1 as $\rho=-1.2606$, $\lambda=0.5783$, $a=1.4338$, $b=11.6641$.
Moreover, just like \cite{Scho}, we take $\alpha=1.75$.
Note that $\alpha$ in \cite{Scho} corresponds to $\alpha-1$ in our setting, and $1-\frac{2\rho}{\calB(T)}$ is greater than 1.75.
In Figures \ref{fig1} and \ref{fig2} below, red crosses and blue circles represent the values of $\xi_t^{(S_T-K)^+}$ and $\Delta_t^{(S_T-K)^+}$, respectively.
We implement the following two types of experiments: First, for fixed strike price $K$, we compute $\xi_t^{(S_T-K)^+}$ and $\Delta_t^{(S_T-K)^+}$ for times $t=0, 0.02,\ldots,0.98$.
Note that we fix $K$ to 900, 1124.47, 1300, which correspond to ``out of the money", ``at the money" and ``in the money", respectively.
Second, $t$ is fixed to 0, 0.5 and 0.9, and we instead vary $K$ from 200 to 2000 at steps of 25, and compute $\xi_t^{(S_T-K)^+}$ and $\Delta_t^{(S_T-K)^+}$.

Now, we discuss implications from Figures \ref{fig1} and \ref{fig2}.
First, $\xi_t^{(S_T-K)^+}$ is always less than or equal to $\Delta_t^{(S_T-K)^+}$.
This fact suggests that local risk-minimization is more risk-averse than the delta-hedge.
Second, Figure \ref{fig1} shows that both $\xi_t^{(S_T-K)^+}$ and $\Delta_t^{(S_T-K)^+}$ are increasing functions of $t$ when the option is ``in the money",
and decreasing when ``at the money" or ``out of the money".
Third, Figure \ref{fig2} implies that both $\xi_t^{(S_T-K)^+}$ and $\Delta_t^{(S_T-K)^+}$ tend to $1$ when the option is ``deep in the money", and $0$ when ``deep out of the money".
In addition, the values of strategies decrease from $1$ to $0$ around ``at the money", and its gradient is steep when the time to maturity is near to $0$.
Finally, the spread between $\xi_t^{(S_T-K)^+}$ and $\Delta_t^{(S_T-K)^+}$ in Figure \ref{fig2} is wider when the option is ``in the money" than ``out of the money".

\begin{figure}[H]
\vspace{-190pt}
  \begin{minipage}[]{\textwidth}
    \centering
    \includegraphics[keepaspectratio,scale=0.25,bb= 0 0 612 792,width=4.0in]{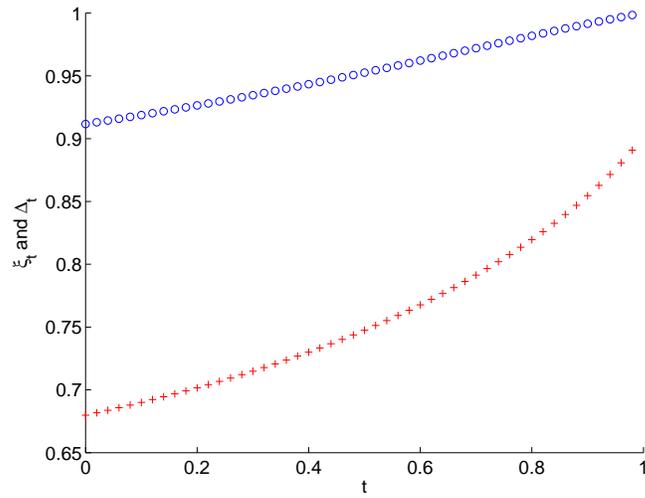}
    \vspace{-95pt}
    \subcaption{Values of $\xi_t^{(S_T-K)^+}$ and $\Delta_t^{(S_T-K)^+}$ when $K$ is fixed to 900 vs. times $t=0, 0.02,\ldots,0.98$.
                In this case, the option is ``in the money" at time $t$.
                Red crosses and blue circles represent the values of $\xi_t^{(S_T-K)^+}$ and $\Delta_t^{(S_T-K)^+}$, respectively.}
    \label{fig11}
  \end{minipage}\\
  \begin{minipage}[]{\linewidth}
    \centering
    \vspace{-95pt}
    \includegraphics[keepaspectratio,scale=0.25,bb = 0 0 612 792,width=4.0in]{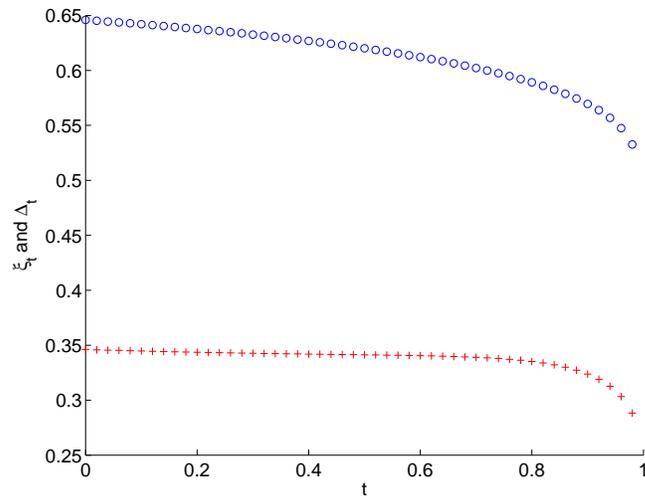}
    \vspace{-95pt}
    \subcaption{Example where the option is ``at the money" at time $t$, that is, $K$ is fixed to 1124.47}\label{fig12}
  \end{minipage}\\
    \begin{minipage}[]{\linewidth}
    \centering
    \vspace{-95pt}
    \includegraphics[keepaspectratio,scale=0.25,bb = 0 0 612 792,width=4.0in]{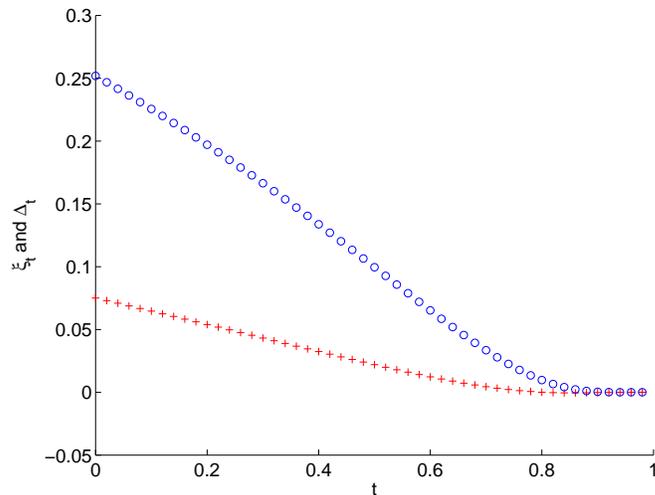}
    \vspace{-95pt}
    \subcaption{Example where $K$ is fixed to 1300, that is, the option is ``out of the money" at time $t$}
    \label{fig13}
  \end{minipage}\\
    \vspace{-9.5pt}
  \caption{Values of $\xi_t^{(S_T-K)^+}$ and $\Delta_t^{(S_T-K)^+}$ for fixed $K$ vs. times $t=0, 0.02,\ldots,0.98$.}
\label{fig1}
\end{figure}

\begin{figure}[H]
\vspace{-190pt}
  \begin{minipage}[]{\textwidth}
    \centering
    \includegraphics[keepaspectratio,scale=0.25,bb= 0 0 612 792,width=4.0in]{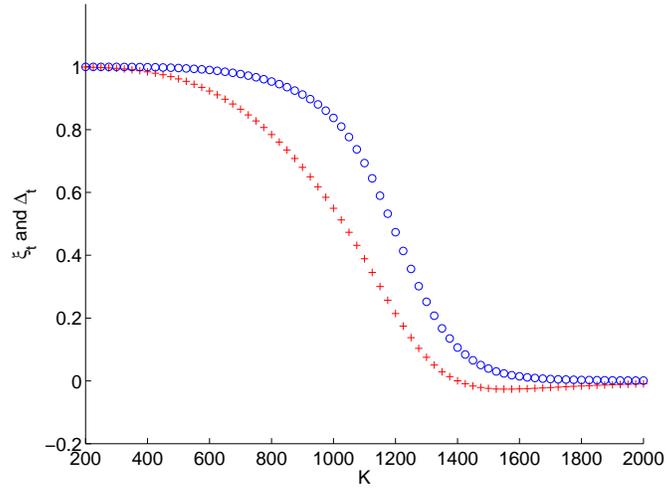}
    \vspace{-95pt}
    \subcaption{Values of $\xi_t^{(S_T-K)^+}$ and $\Delta_t^{(S_T-K)^+}$ at $t=0$ vs. strike price $K$ from 200 to 2000 at steps of 25.
                Red crosses and blue circles represent the values of $\xi_t^{(S_T-K)^+}$ and $\Delta_t^{(S_T-K)^+}$, respectively.}\label{fig21}
  \end{minipage}\\
  \begin{minipage}[]{\linewidth}
    \centering
    \vspace{-95pt}
    \includegraphics[keepaspectratio,scale=0.25,bb= 0 0 612 792,width=4.0in]{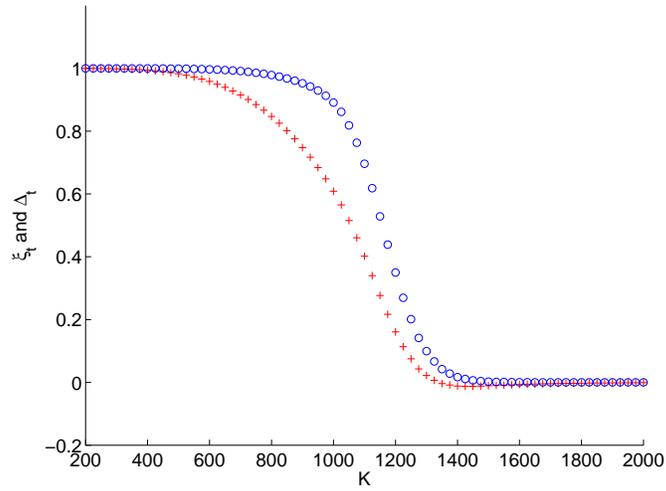}
    \vspace{-95pt}
    \subcaption{Example where $t=0.5$.}\label{fig22}
  \end{minipage}\\
    \begin{minipage}[]{\linewidth}
    \centering
    \vspace{-95pt}
    \includegraphics[keepaspectratio,scale=0.25,bb= 0 0 612 792,width=4.0in]{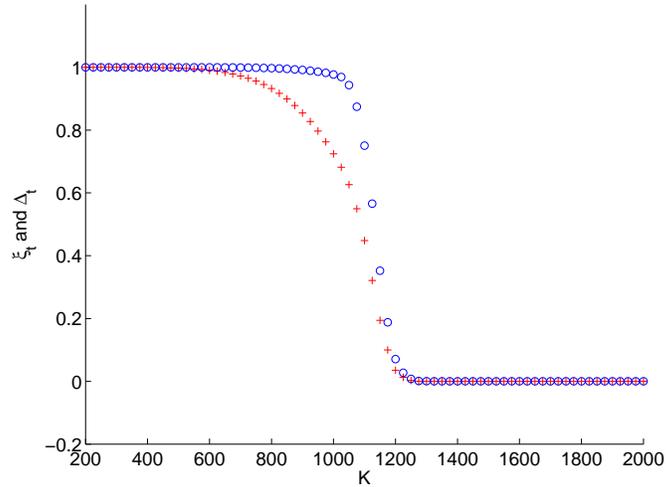}
    \vspace{-95pt}
    \subcaption{Example where $t=0.9$.}\label{fig23}
  \end{minipage}\\
    \vspace{-9.5pt}
  \caption{Values of $\xi_t^{(S_T-K)^+}$ and $\Delta_t^{(S_T-K)^+}$ at fixed $t$ vs. strike price $K$ from 200 to 2000 at steps of 25.}
\label{fig2}
\end{figure}

\setcounter{equation}{0}
\section{Conclusions}
We obtain explicit representations of LRM strategies of call and put options for the BNS models given by (\ref{SDE-sigma}) and (\ref{eq-S}), and implement related numerical experiments.
We impose only Assumption \ref{ass1} as the standing assumptions.
Recall that Assumption \ref{ass1} does not exclude the two important examples: IG-OU and Gamma-OU, although parameters are restricted.
Our discussion is based on the framework of \cite{AS}.
We confirm all the additional conditions imposed in \cite{AS}.
Above all, we need some integrability conditions on the underlying contingent claim $F$.
For example, we need $Z_TF\in L^2(\bbP)$, which is almost equivalent to $Z_TS_T\in L^2(\bbP)$ if $F$ is a call option.
However, $Z_TS_T$ is not in $L^2(\bbP)$ in our setting, which means that an additional condition is needed to treat call options directly in the framework of \cite{AS}.
Thus, we consider put options first in this paper, as they are bounded. LRM strategies for call options are then given as a corollary.
With this simple idea, we do not need to impose any additional condition.

Moreover, to demonstrate condition C4, we need to investigate the Malliavin differentiability of the process $Z$.
Note that $Z$ is a solution to the SDE (\ref{SDE-Z}). \cite{NOP} showed the Malliavin differentiability of solutions to Markovian--type SDEs with the Lipschitz condition.
However, the SDE (\ref{SDE-Z}) is not Markovian, because $u_s$ and $\theta_{s,x}$ are random.
In Section 5, as an extension of Section 17 in \cite{NOP}, we show that $Z_t\in\bbD^{1,2}$.
This result should be a valuable mathematical contribution in its own right.
Recall that $u_s$ and $\theta_{s,x}$ are bounded by Lemma \ref{lem-ut-1},
and the Malliavin derivatives of $u_s$ and $\theta_{s,x}$ are equivalent to $O(1/z)$ and $O(1/\sqrt{z})$ simultaneously by Lemmas \ref{lem-ut-2} and \ref{lem-ut-3}.
These facts play a vital role in the demonstration of the Malliavin differentiability of $Z$.

We consider, throughout the paper, BNS models for which the asset price process is given by (\ref{eq-S}).
Actually, the general form of BNS models is as follows:
\[
S_t=S_0\exp\l\{\int_0^t(\mu+\beta\sigma_s^2)ds+\int_0^t\sigma_sdW_s+\rho J_t\r\},
\]
where the parameter $\beta\in\bbR$ is called the volatility risk premium.
In other words, we restrict $\beta$ to $-1/2$.
If $\beta\neq-1/2$, the boundedness of $u_s$ and $\theta_{s,x}$ no longer holds, from which it is not easy to show that $Z_T\in\bbD^{1,2}$.
Thus, formulating a Malliavin calculus under the MMM, \cite{A} took a different approach to study the case of $\beta\in\bbR$ and $\rho=0$.
On the other hand, some new ideas are needed to treat fully general case.
It remains for future research.

\appendix
\setcounter{equation}{0}
\section{Appendix}
\renewcommand{\theequation}{A.\arabic{equation}}
\setcounter{equation}{0}

\subsection{Theorem 3.7 of \cite{AS}}

Theorem 3.7 of \cite{AS}, which provides an explicit representation formula of LRM strategies for L\'evy markets, is frequently referred to in this paper.
Therefore, we introduce its statement for BNS models under Assumption \ref{ass1}.
Note that, although Assumption 2.1 of \cite{AS} is imposed on Theorem 3.7 of \cite{AS}, it is satisfied under Assumption \ref{ass1}.
For more details, see Remark \ref{rem1}.

\begin{thm}[Theorem 3.7 of \cite{AS}]
\label{thm-AS37}
Let $F$ be an $L^2(\bbP)$ random variable satisfying the following three conditions:
\begin{description}
\item[AS1] (Assumption 2.6 in \cite{AS}) $Z_TF$ is in $L^2(\bbP)$.
\item[AS2] (Assumption 3.4 in \cite{AS}) Conditions (C1)--(C6) for $F$ are satisfied.
\item[AS3] ((3.1) in \cite{AS}) We have
\[
\bbE\l[\int_0^T\l\{(h^0_t)^2+\int_0^\infty(h^1_{t,z})^2\nu(dz)\r\}dt\r]<\infty,
\]
where $h_{t,z}^1=\bbE_{\tP}[F(H^*_{t,z}-1)+zH^*_{t,z}D_{t,z}F|\calF_{t-}]$ and
\[
h_t^0=\bbE_{\tP}\l[D_{t,0}F-F\l[\int_0^TD_{t,0}u_sdW^{\tP}_s+\int_0^T\int_{\bbR\backslash\{0\}}\frac{D_{t,0}\theta_{s,x}}{1-\theta_{s,x}}\tN^{\tP}(ds,dx)\r]\Big|\calF_{t-}\r].
\]
\end{description}
Then, the LRM strategy $\xi^F$ for claim $F$ is given by
\[
\xi^F_t=\frac{1}{S_{t-}(\sigma^2_t+C_\rho)}\l\{h^0_t\sigma_t+\int_0^\infty h^1_{t,z}(e^{\rho z}-1)\nu(dz)\r\}.
\]
\end{thm}

\subsection{Properties of $\sigma_t$, and related Malliavin derivatives}

The squared volatility process $\sigma_t^2$, given as a solution to the SDE (\ref{SDE-sigma}), is represented as
\begin{equation}
\label{eq-sigma-0}
\sigma_t^2=e^{-\lambda t}\sigma_0^2+\int_0^te^{-\lambda(t-s)}dJ_s.
\end{equation}
Remark that we have
\begin{equation}
\label{eq-sigma-1}
\sigma_t^2\geq e^{-\lambda t}\sigma_0^2\geq e^{-\lambda T}\sigma_0^2,
\end{equation}
and
\begin{equation}
\label{eq-sigma-2}
\int_0^t\sigma_s^2ds
=    \frac{1}{\lambda}(J_t-\sigma_t^2+\sigma_0^2)
\leq \frac{1}{\lambda}(J_t+\sigma_0^2).
\end{equation}

Next, we calculate some related Malliavin derivatives.

\begin{lem}
\label{lem-sigma-1}
For any $s\in[0,T]$, we have $\sigma_s^2\in\bbD^{1,2}$; and
\begin{equation}
\label{eq-sigma-3}
D_{t,z}\sigma_s^2=e^{-\lambda(s-t)}{\bf 1}_{[0,s]\times(0,\infty)}(t,z)
\end{equation}
for $(t,z)\in[0,T]\times[0,\infty)$.
\end{lem}

\proof
We can rewrite (\ref{eq-sigma-0}) as
\begin{align*}
\sigma_s^2
&= e^{-\lambda s}\sigma_0^2+\int_0^s\int_0^\infty e^{-\lambda(s-u)}x\nu(dx)du \\
&  \hspace{7mm}+\int_{[0,T]\times[0,\infty)}e^{-\lambda(s-u)}{\bf 1}_{[0,s]\times(0,\infty)}(u,x)Q(du,dx).
\end{align*}
Moreover, we have $\int_{[0,T]\times[0,\infty)}e^{-2\lambda(s-u)}{\bf 1}_{[0,s]\times(0,\infty)}(u,x)q(du,dx)<\infty$.
By Definition \ref{def-Malliavin}, the lemma follows.
\fin

\begin{lem}
\label{lem-sigma-2}
For any $s\in[0,T]$, we have $\sigma_s\in\bbD^{1,2}$; and
\[
D_{t,z}\sigma_s=\frac{\sqrt{\sigma_s^2+ze^{-\lambda(s-t)}}-\sigma_s}{z}{\bf 1}_{[0,s]\times(0,\infty)}(t,z)
\]
for $(t,z)\in[0,T]\times[0,\infty)$.
Furthermore, we have $0\leq D_{t,z}\sigma_s\leq\frac{1}{\sqrt{z}}{\bf 1}_{[0,s]}(t)$ for $z>0$.
\end{lem}

\proof
Taking a $C^1$-function $f$ such that $f^\prime$ is bounded; and $f(r)=\sqrt{r}$ for $r\geq e^{-\lambda T}\sigma_0^2$, we have $\sigma_s=f(\sigma_s^2)$ by (\ref{eq-sigma-1}).
Proposition 2.6 in \cite{Suz} implies $\sigma_s\in\bbD^{1,2}$, $D_{t,0}\sigma_s=f^\prime(\sigma_s^2)D_{t,0}\sigma_s^2=0$; and
\[
D_{t,z}\sigma_s
= \frac{f(\sigma_s^2+zD_{t,z}\sigma_s^2)-f(\sigma_s^2)}{z}
= \frac{\sqrt{\sigma_s^2+ze^{-\lambda(s-t)}}-\sigma_s}{z}{\bf 1}_{[0,s]}(t)
\]
for $z>0$, as $D_{t,z}\sigma_s^2$ is nonnegative by (\ref{eq-sigma-3}).
In addition, we have $D_{t,z}\sigma_s\leq\frac{\sqrt{ze^{-\lambda(s-t)}}}{z}{\bf 1}_{[0,s]}(t) \leq \frac{1}{\sqrt{z}}{\bf 1}_{[0,s]}(t)$ for $z>0$.
\fin

\begin{lem}
\label{lem-sigma-3}
We have $\int_0^T\sigma_s^2ds\in\bbD^{1,2}$; and
\[
D_{t,z}\int_0^T\sigma_s^2ds=\calB(T-t){\bf 1}_{(0,\infty)}(z)
\]
for $(t,z)\in[0,T]\times[0,\infty)$, where the function $\calB$ is defined in Assumption \ref{ass1}.
\end{lem}

\proof
First, we have
\begin{align*}
\int_0^T\sigma_s^2ds
&= \sigma_0^2\int_0^Te^{-\lambda s}ds+\int_0^T\int_0^se^{-\lambda(s-u)}dJ_uds \\
&= \sigma_0^2\frac{1-e^{-\lambda T}}{\lambda}+\int_0^T\int_u^Te^{-\lambda(s-u)}dsdJ_u
   =\sigma_0^2\calB(T)+\int_0^T\calB(T-u)dJ_u.
\end{align*}
From the view of Definition \ref{def-Malliavin},
we obtain $\int_0^T\sigma_s^2ds\in\bbD^{1,2}$ and $D_{t,z}\int_0^T\sigma_s^2ds= \calB(T-t){\bf 1}_{(0,\infty)}(z)$ for $(t,z)\in[0,T]\times(0,\infty)$.
\fin

\begin{lem}
\label{lem-sigma-4}
We have $\int_0^T\sigma_sdW_s\in\bbD^{1,2}$ and
\[
D_{t,z}\int_0^T\sigma_sdW_s=\sigma_t{\bf 1}_{\{0\}}(z)+\int_t^T\frac{\sqrt{\sigma_s^2+ze^{-\lambda(s-t)}}-\sigma_s}{z}dW_s{\bf 1}_{(0,\infty)}(z).
\]
for $(t,z)\in[0,T]\times[0,\infty)$.
\end{lem}

\proof
To begin, we show $\sigma\in\bbL_0^{1,2}$. Lemma \ref{lem-sigma-2} implies $\sigma_s\in\bbD^{1,2}$ for any $s\in[0,T]$. We have $\bbE\l[\int_0^T\sigma_s^2ds\r]<\infty$ by (\ref{eq-sigma-2}) and the integrability of $J_T$. As $|D_{t,z}\sigma_s|^2\le\frac{1}{z}$ by Lemma \ref{lem-sigma-2}, item (c) of the definition of $\bbL_0^{1,2}$ is satisfied. Hence, Lemma 3.3 in \cite{DI} provides $\int_0^T\sigma_sdW_s\in\bbD^{1,2}$ and
\begin{align*}
D_{t,z}\int_0^T\sigma_sdW_s
&= D_{t,z}\int_{[0,T]\times[0,\infty)}\sigma_s\cdot{\bf 1}_{\{0\}}(x)Q(ds,dx)
   = \sigma_t{\bf 1}_{\{0\}}(z)+\int_0^TD_{t,z}\sigma_sdW_s \\
&= \sigma_t{\bf 1}_{\{0\}}(z)+\int_t^T\frac{\sqrt{\sigma_s^2+ze^{-\lambda(s-t)}}-\sigma_s}{z}dW_s{\bf 1}_{(0,\infty)}(z)
\end{align*}
for $(t,z)\in [0,T]\times[0,\infty)$ by Lemma \ref{lem-sigma-2}.
\fin

Lastly, we calculate $D_{t,z}L_T$ as follows:

\begin{prop}
\label{prop-LT-1}
$L_T\in\bbD^{1,2}$ and, for $(t,z)\in[0,T]\times[0,\infty)$, we have
\[
D_{t,z}L_T=\sigma_t{\bf 1}_{\{0\}}(z)+\l\{-\frac{1}{2}\calB(T-t)+\int_t^T\frac{\sqrt{\sigma_s^2+ze^{-\lambda(s-t)}}-\sigma_s}{z}dW_s+\rho\r\}{\bf 1}_{(0,\infty)}(z).
\]
\end{prop}

\proof
By (\ref{eq-L}), we have $L_T=\mu T-\frac{1}{2}\int_0^T\sigma_s^2ds+\int_0^T\sigma_sdW_s+\rho J_T$.
Because $J_T\in\bbD^{1,2}$ and $D_{t,z}J_T={\bf 1}_{(0,\infty)}(z)$, we obtain this proposition by Lemmas \ref{lem-sigma-3} and \ref{lem-sigma-4}.
\fin

\subsection{Properties of $u_s$ and $\theta_{s,x}$, and related Malliavin derivatives}

We begin with the definition of two constants as follows:
\begin{equation}
\label{eq-def-C}
C_u     :=\max\l\{\frac{|\alpha|e^{\frac{\lambda T}{2}}}{\sigma_0},\frac{|\alpha|}{C_\rho}\r\};
\mbox{ \ \ and \ \ }
C_\theta:=\max\l\{\frac{|\alpha|}{C_\rho},1\r\}.
\end{equation}
The next lemma is cited often throughout the paper.

\begin{lem}
\label{lem-ut-1}
For any $s\in[0,T]$ and any $x\in(0,\infty)$, the following hold:
\begin{enumerate}
\item $|u_s|\leq C_u$,
\item $|\theta_{s,x}|\leq C_\theta$; and $|\theta_{s,x}|\leq C_\theta(1-e^{\rho x})\leq C_\theta|\rho|x$,
\item $\theta_{s,x}<1-e^{\rho x}$,
\item $|\log(1-\theta_{s,x})|\leq C_\theta|\rho|x$,
\item $\frac{1}{1-\theta_{s,x}}<\hat{C}_\theta$ for some $\hat{C}_\theta>0$.
\end{enumerate}
\end{lem}

\proof
\begin{enumerate}
\item We have $|u_s|\leq\frac{|\alpha|}{\sigma_s}\leq\frac{|\alpha|e^{\frac{\lambda T}{2}}}{\sigma_0}$ for any $s\in[0,T]$ by (\ref{eq-sigma-1}).
\item $|\theta_{s,x}|\leq\frac{|\alpha|}{C_\rho}(1-e^{\rho x})\leq C_\theta$ and $1-e^{\rho x}\leq|\rho|x$ for any $x>0$.
\item As seen in Remark \ref{rem1}, $\frac{\alpha}{\sigma^2_s+C_\rho}>-1$ for any $s\in[0,T]$. We have then $\theta_{s,x}<1-e^{\rho x}$.
\item When $\theta_{s,x}\geq 0$, we have $0\geq\log(1-\theta_{s,x}) >\log(1-(1-e^{\rho x}))=\rho x\geq C_\theta\rho x$.
      On the other hand, if $\theta_{s,x}< 0$, then $0<\log(1-\theta_{s,x})\leq -\theta_{s,x}\leq C_\theta|\rho|x$.
\item If $\theta_{s,x}\leq0$, then $\frac{1}{1-\theta_{s,x}}\leq1$; otherwise, if $\theta_{s,x}>0$, equivalently $\alpha<0$,
      then $1-\theta_{s,x} = 1+\frac{\alpha}{\sigma^2_s+C_\rho}(1-e^{\rho x}) \geq 1+\frac{\alpha}{\sigma^2_s+C_\rho} \geq 1+\frac{\alpha}{e^{-\lambda T}\sigma^2_0+C_\rho}>0$
      by Assumption \ref{ass1}.
      This completes the proof.
\end{enumerate}
\fin

Next, we calculate some Malliavin derivatives related to $u_s$ and $\theta_{s,x}$.

\begin{lem}
\label{lem-ut-2}
For any $s\in[0,T]$, we have $u_s\in\bbD^{1,2}$; and
\begin{align}
D_{t,z}u_s
&= \frac{f_u(\sigma_s+zD_{t,z}\sigma_s)-f_u(\sigma_s)}{z}{\bf 1}_{[0,s]\times(0,\infty)}(t,z) \nonumber \\
&= \frac{f_u\l(\sqrt{\sigma^2_s+ze^{-\lambda(s-t)}}\r)-f_u(\sigma_s)}{z}{\bf 1}_{[0,s]\times(0,\infty)}(t,z)
\label{eq-ut-1}
\end{align}
for $(t,z)\in[0,T]\times[0,\infty)$, where $f_u(r):=\frac{\alpha r} {r^2+C_\rho}$ for $r\in\bbR$. Moreover, we have
\[
|D_{t,z}u_s|\leq\frac{C_u}{\sqrt{z}}{\bf 1}_{[0,s]}(t) \quad\mbox{and}\quad |D_{t,z}u_s|\leq\frac{C_u^\prime}{z}{\bf 1}_{[0,s]}(t)
\]
for some $C_u^\prime>0$.
\end{lem}

\proof
Note that $f^\prime_u(r)=\alpha\frac{C_\rho-r^2}{(r^2+C_\rho)^2}$ and $|f^\prime_u(r)|\leq \frac{|\alpha|}{C_\rho}\leq C_u$.
Because $u_s=f_u(\sigma_s)$ and $\sigma_s\in\bbD^{1,2}$, Proposition 2.6 in \cite{Suz}, together with Lemma \ref{lem-sigma-2}, implies $u_s\in\bbD^{1,2}$ and (\ref{eq-ut-1}).
In particular, we have $D_{t,0}u_s = f^\prime_u(\sigma_s)D_{t,0}\sigma_s = 0$.
Further, Lemma \ref{lem-sigma-2} again yields $|D_{t,z}u_s|\leq\frac{1}{z}|zD_{t,z}\sigma_s|C_u\leq\frac{1}{\sqrt{z}}{\bf 1}_{[0,s]}(t)C_u$.
Moreover, as $f_u(r)$ is bounded, we can find a $C_u^\prime>0$ such that $|D_{t,z}u_s|\leq\frac{C_u^\prime}{z}$.
\fin

\begin{lem}
\label{lem-ut-3}
For any $(s,x)\in[0,T]\times(0,\infty)$, we have $\theta_{s,x}\in\bbD^{1,2}$; and
\begin{align}
D_{t,z}\theta_{s,x}
&= \frac{f_\theta(\sigma_s+zD_{t,z}\sigma_s)-f_\theta(\sigma_s)}{z}(e^{\rho x}-1){\bf 1}_{[0,s]\times(0,\infty)}(t,z) \nonumber \\
&= \frac{f_\theta\l(\sqrt{\sigma^2_s+ze^{-\lambda(s-t)}}\r)-f_\theta(\sigma_s)}{z}(e^{\rho x}-1){\bf 1}_{[0,s]\times(0,\infty)}(t,z)
\label{eq-ut-3}
\end{align}
for $(t,z)\in[0,T]\times[0,\infty)$, where $f_\theta(r):=\frac{\alpha}{r^2+C_\rho}$ for $r\in\bbR$.
Moreover, we have
\begin{equation}
\label{eq-ut-3-2}
|D_{t,z}\theta_{s,x}|\leq\frac{C^\prime_\theta}{\sqrt{z}}(1-e^{\rho x}){\bf 1}_{[0,s]}(t)
\mbox{ \ \ and \ \ }
|D_{t,z}\theta_{s,x}|\leq\frac{2C_\theta}{z}(1-e^{\rho x}){\bf 1}_{[0,s]}(t)
\end{equation}
for some $C_\theta^\prime>0$.
\end{lem}

\proof
Note that $\theta_{s,x}=f_\theta(\sigma_s)(e^{\rho x}-1)$ and $f^\prime_\theta(r)=-\frac{2\alpha r}{(r^2+C_\rho)^2}$.
Hence, $|f^\prime_\theta(r)|$ is bounded. Therefore, the same argument as Lemma \ref{lem-ut-2} implies (\ref{eq-ut-3}).
In addition, (\ref{eq-ut-3-2}) is given by the boundedness of $f_\theta$ and $f_\theta^\prime$.
\fin

\begin{lem}
\label{lem-ut-4}
For any $(s,x)\in[0,T]\times(0,\infty)$, we have $\log(1-\theta_{s,x})\in \bbD^{1,2}$; and
\[
D_{t,z}\log(1-\theta_{s,x})=\frac{\log(1-\theta_{s,x}-zD_{t,z}\theta_{s,x})-\log(1-\theta_{s,x})}{z}{\bf 1}_{(0,\infty)}(z)
\]
for $(t,z)\in[0,T]\times[0,\infty)$.
Moreover, we have $|D_{t,z}\log(1-\theta_{s,x})|\leq|D_{t,z}\theta_{s,x}|e^{-\rho x}$.
\end{lem}

\proof
For $x>0$, we denote
\begin{align*}
g_x(r):=\l\{
        \begin{array}{ll}
        \log(1-r),                          & r< 1-e^{\rho x},   \\
        -e^{-\rho x}r+e^{-\rho x}-1+\rho x, & r\geq 1-e^{\rho x}.
        \end{array}\r.
\end{align*}
Note that $g_x$ is a $C^1$-function satisfying $|g^\prime_x(r)|\leq e^{-\rho x}$ for all $r\in\bbR$.
Because $\theta_{s,x}\in\bbD^{1,2}$ and $\log(1-\theta_{s,x})=g_x(\theta_{s,x})$ by item 3 of Lemma \ref{lem-ut-1}, we have
\[
D_{t,z}\log(1-\theta_{s,x})=\frac{g_x(\theta_{s,x}+zD_{t,z}\theta_{s,x})-g_x(\theta_{s,x})}{z}{\bf 1}_{(0,\infty)}(z).
\]
Lemma \ref{lem-ut-3} implies, for $t\in[0,s]$ and $z\in(0,\infty)$,
\begin{align}
\theta_{s,x}+zD_{t,z}\theta_{s,x}
&= f_\theta\l(\sqrt{\sigma^2+ze^{-\lambda(s-t)}}\r)(e^{\rho x}-1) \nonumber \\
&= \frac{\alpha(e^{\rho x}-1)}{\sigma^2_s+ze^{-\lambda(s-t)}+C_\rho}
   < 1-e^{\rho x}.
\label{eq-ut-5}
\end{align}
We have then $g_x(\theta_{s,x}+zD_{t,z}\theta_{s,x})=\log(1-\theta_{s,x}-zD_{t,z}\theta_{s,x})$.
\fin

\subsection{On $D_{t,z}\log Z_T$}

We show $\log Z_T\in\bbD^{1,2}$ and calculate $D_{t,z}\log Z_T$.
(\ref{eq-MMM-1}) implies that
\begin{align}
\log Z_T
&= -\int_0^Tu_sdW_s-\frac{1}{2}\int_0^Tu_s^2ds+\int_0^T\int_0^\infty\log(1-\theta_{s,x})\tN(ds,dx) \nonumber \\
&  \hspace{5mm}+\int_0^T\int_0^\infty[\log(1-\theta_{s,x})+\theta_{s,x}]\nu(dx)ds.
\label{eq-logZT-1}
\end{align}
We discuss each term of (\ref{eq-logZT-1}) separately.
As seen in Section \ref{sect-proof}, we have $u\in\bbL_0^{1,2}$.
Therefore, Lemma 3.3 of \cite{DI} implies that $D_{t,0}\int_0^Tu_sdW_s=u_t+\int_0^TD_{t,0}u_sdW_s=u_t$, and $D_{t,z}\int_0^Tu_sdW_s=\int_0^TD_{t,z}u_sdW_s$ for $z>0$.
Similarly, we have
$D_{t,0}\int_0^T\int_0^\infty\log(1-\theta_{s,x})\tN(ds,dx)=0$, and
\[
D_{t,z}\int_0^T\int_0^\infty\log(1-\theta_{s,x})\tN(ds,dx)=\frac{\log(1-\theta_{t,z})}{z}+\int_0^T\int_0^\infty D_{t,z}\log(1-\theta_{s,x})\tN(ds,dx)
\]
for $z>0$.
As for $D_{t,z}\int_0^Tu_s^2ds$, because $u^2\in\bbL_0^{1,2}$ by Section \ref{sect-proof}, Lemma 3.2 of \cite{DI} yields
\[
D_{t,z}\int_0^Tu_s^2ds=\int_0^TD_{t,z}u_s^2ds=2\int_0^Tu_sD_{t,z}u_sds+z\int_0^T(D_{t,z}u_s)^2ds
\]
for $z\geq0$.
In particular, $D_{t,0}\int_0^Tu_s^2ds=0$. For the fourth term of (\ref{eq-logZT-1}), because $\log(1-\theta)+\theta\in\widetilde{\bbL}_1^{1,2}$, Proposition 3.5 of \cite{Suz} implies
\[
D_{t,z}\int_0^T\int_0^\infty[\log (1-\theta_{s,x})+\theta_{s,x}]\nu(dx)ds=\int_0^T\int_0^\infty[D_{t,z}\log(1-\theta_{s,x})+D_{t,z}\theta_{s,x}]\nu(dx)ds
\]
for $z\geq 0$. Collectively, we conclude the following:

\begin{prop}
\label{prop-logZT}
We have $\log Z_T\in\bbD^{1,2}$, $D_{t,0}\log Z_T=u_t$; and
\begin{align*}
D_{t,z}\log Z_T
&= -\int_0^TD_{t,z}u_sdW_s-\int_0^Tu_sD_{t,z}u_sds-\frac{z}{2}\int_0^T(D_{t,z}u_s)^2ds \\
&  \hspace{5mm}+\int_0^T\int_0^\infty D_{t,z}\log(1-\theta_{s,x})\tN(ds,dx) \\
&  \hspace{5mm}+\int_0^T\int_0^\infty[D_{t,z}\log(1-\theta_{s,x})+D_{t,z}\theta_{s,x}]\nu(dx)ds+\frac{\log(1-\theta_{t,z})}{z}
\end{align*}
for $z>0$.
\end{prop}

\begin{center}
{\bf Acknowledgments}
\end{center}
Takuji Arai gratefully acknowledges the financial support of Ishii Memorial Securities Research Promotion Foundation, and
Scientific Research (C) No.15K04936 from the Ministry of Education, Culture, Sports, Science and Technology of Japan.



\begin{thebibliography}{00}
\bibitem{A}
Arai, T.: Local risk-minimization for Barndorff-Nielsen and Shephard models with volatility risk premium, to appear in Advances in Mathematical Economics (2015)
\bibitem{AIS}
Arai, T., Imai, Y., Suzuki, R.: Numerical analysis on local risk-minimization for exponential Levy models,
to appear in International Journal of Theoretical and Applied Finance (2015)
\bibitem{AS}
Arai, T., Suzuki, R.: Local risk minimization for L\'evy markets. International Journal of Financial Engineering, 2, 1550015 (2015)
\bibitem{BNS1}
Barndorff-Nielsen, O.E., Shephard, N.: Modelling by L\'evy processes for financial econometrics. In: Barndorff-Nielsen, O.E., Mikosch,T., Resnick, S. (eds.):
L\'evy processes---Theory and Applications, pp. 283--318.
Birkh\"auser, Basel (2001)
\bibitem{BNS2}
Barndorff-Nielsen, O.E., Shephard, N.: Non-Gaussian Ornstein-Uhlenbeck based models and some of their uses in financial econometrics. J.R. Statistic. Soc. 63, 167--241 (2001)
\bibitem{BD}
Benth, F.E., Detering, N.: Pricing and Hedging Asian-Style Options in Energy. Finance and Stochastics,19, 849--889 (2015)
\bibitem{CM}
Carr, P., D. Madan: Option valuation using the fast Fourier transform. Journal of Computational Finance, 2, 61--73 (1999)
\bibitem{CT}
Cont, R., Tankov P.: Financial Modelling with Jump Processes. Chapman \& Hall, London (2004)
\bibitem{CTV}
Cont, R., Tankov, P., Voltchkova, E.: Hedging with options in models with jumps. In: F. Benth et al. (Ed.), Stochastic analysis and applications.
The Abel symposium 2005 (pp.197-217). Berlin: Springer (2007)
\bibitem{DI}
Delong, L., Imkeller, P.: On Malliavin's differentiability of BSDEs with time delayed generators driven by Brownian motions and Poisson random measures.
Stochastic Process. Appl. 120, 1748--1775 (2010)
\bibitem{NOP}
Di Nunno, G., \O ksendal, B., Proske, F.: Malliavin Calculus for L\'evy Processes with Applications to Finance. Springer, Berlin (2009)
\bibitem{I}
Ishikawa, Y.: Stochastic Calculus of Variations for Jump Processes. Walter De Gruyter, Berlin (2013)
\bibitem{KP}
Kallsen, J., Pauwels, A.: Variance-optimal hedging in general affine stochastic volatility models. Adv. Appl. Prob. 42, 83--105 (2010)
\bibitem{KV}
Kallsen, J., Vierthauer, R.: Quadratic hedging in affine stochastic volatility models. Rev. Deriv. Res. 12, 3--27 (2009)
\bibitem{NV}
Nicolato, E., Venardos, E.: Option Pricing in Stochastic Volatility Models of the Ornstein-Uhlenbeck type. Mathematical Finance. 13 (4), 445--466 (2003)
\bibitem{P}
Protter, P.: Stochastic Integration and Differential Equations. Springer, Berlin (2004)
\bibitem{Scho}
Schoutens, W.: L\'evy Processes in Finance: Pricing Financial Derivatives. John Wiley \& Sons, Hoboken (2003)
\bibitem{Sch}
Schweizer, M.: A Guided Tour through Quadratic Hedging Approaches. In: Jouini, E., Cvitani{\'c}, J., Musiela, M. (eds.):
Option Pricing, Interest Rates and Risk Management (Handbooks in Mathematical Finance), pp. 538--574. Cambridge University Press, Cambridge (2001)
\bibitem{Sch3}
Schweizer, M.: Local Risk-Minimization for Multidimensional Assets and Payment Streams. Banach Center Publ. 83, 213--229 (2008)
\bibitem{Situ}
Situ, R.: Theory of Stochastic Differential Equations with Jumps and Applications (Mathematical and Analytical Techniques with Applications to Engineering). Springer, Berlin (2005)
\bibitem{Suz}
Suzuki, R.: A Clark-Ocone type formula under change of measure for L\'evy processes with $L^2$-L\'evy measure. Commun. Stoch. Anal. 7, 383--407 (2013)
\bibitem{S07}
Sol\'e, J.L., Utzet, F., Vives, J.: Canonical L\'evy process and Malliavin calculus. Stochastic Process. Appl. 117, 165--187 (2007)
\end{thebibliography}
\end{document}